\documentclass[twocolumn]{aastex631}
\usepackage{amsmath}

\newcommand{\logHa}{log$_{10}$({\rm H}$\alpha$\,{\rm SB})}
\def\objects{94}

\begin{document}

%\title{Hosts of Ia Supernovae Survey I: Local Host-Galaxy H$\alpha$ Surface Brightness and the Hubble Residuals of Type Ia Supernovae}

\title{North+Lone Star Supernova Host Survey I: Local Host-Galaxy H$\alpha$ Surface Brightness and the Hubble Residuals of Type Ia Supernovae}

\author[0009-0006-9538-4781]{Ann M. Isaacs} 
\email{isaac413@umn.edu}
\affiliation{School of Physics and Astronomy, University of Minnesota, 116 Church Street SE, Minneapolis, MN 55455, US}
\author[0000-0003-3142-997X]{Patrick Kelly} 
\affiliation{School of Physics and Astronomy, University of Minnesota, 116 Church Street SE, Minneapolis, MN 55455, US}
\author[0000-0003-1349-6538]{J. Craig Wheeler} 
\affiliation{Department of Astronomy, University of Texas at Austin, Austin, TX, USA}

\date{\today}

\begin{abstract}
We present optical integral-field unit (IFU) spectroscopy acquired with the George and Cynthia Mitchell Spectrograph on the Harlan J. Smith telescope at McDonald Observatory of \objects\ galaxies ($0.01<z<0.058$) that have hosted Type Ia supernovae (SNe Ia). We selected host galaxies with star-forming morphology, consistent with the criteria used by \citet{riess2022}. We measured the H$\alpha$ surface brightness of each host galaxy within 1 kpc of the location of the supernova. Using distances from the Pantheon+ sample, we find a step in Hubble residuals compared to local H$\alpha$ surface brightness of $-0.097 \pm 0.051$ mag at 1.9$\sigma$ significance in a sample of 73 host galaxies, where SNe in environments with smaller H$\alpha$ surface brightness are, on average, less luminous after correction for light-curve shape and color. Almost all of the SNe in our sample were discovered by targeted surveys. Using an independent sample primarily from the untargeted Nearby Supernova Factory survey, \citet{rigault2020} found a step of 0.045 $\pm$ 0.029 mag where SNe in passive environments are instead brighter, which is in 2.4$\sigma$ tension with our measurement. \citet{rigault2013} designated SNe Ia comparatively small HRs ($<-0.1$) and faint local H$\alpha$ surface brightness (SB) ($<$log$_{10}$(H$\alpha$\,SB / (erg$^{-1}$ s$^{-1}$ kpc$^2$))=38.32 as the $M_2$ population. SNe that would be classified as $M_2$ are less highly represented in our sample (7\% versus 21\%). When we include an additional twelve early-type galaxies, the number of $M_2$ SNe is almost doubled, although the tension with the HR step measured by \citet{rigault2020} persists at 1.7$\sigma$. 
\end{abstract} 

\section{Introduction} 
\label{sec:intro}
Type Ia supernovae (SNe Ia) are one of the most precise tools for measuring the expansion history of the universe, and enabled the discovery of the accelerating expansion of the universe \citep{riess1998, permutter1999}.
After correction for light curve decline rate and color, \citep[e.g.,][]{phillips1993, riess1996, tripp1998}, they can be used to measure distances in the universe to within $\sim$8\%. Measurements of the Hubble constant $H_0$ using SNe Ia, however, are in tension with results from the Planck survey of the cosmic microwave background (CMB), with local SNe Ia measurements by the SH0ES team giving a value of $73.04\pm1.04$ km s$^{-1}$ Mpc$^{-1}$\citep{riess2022} and measurements of the CMB yielding $67.4\pm0.5$ km s$^{-1}$ Mpc$^{-1}$ \citep{planck2018}. The tension between these two measurements stands at nearly 5$\sigma$. To improve distance measurements with SNe Ia, many efforts have examined potential correlations between SN Ia with calibrated luminosities and the properties of their host galaxies, including dependence on their stellar masses \citep[e.g.,][]{kelly2010, lampeitl2010, sullivan2010, smith2020, kelsey2021}. Additional potential connections have been investigated as well, especially with host-galaxy star formation rate (SFR). \cite{rigault2013} use a measurement of the local SFR (LSFR) estimated from H$\alpha$ surface brightness and find that SNe Ia in passive environments are brighter after light curve correction than those in star-forming environments. In this paper, we search for this relation in an almost entirely new sample of SNe Ia host galaxies by comparing their local H$\alpha$ surface brightnesses with their Hubble residuals (HRs). We define the HR here as the difference between the distance modulus inferred from the SNe Ia light curve and that predicted by the Hubble relation, $\mu_{SN}-\mu_z$, where $\mu_{SN}$ is the distance modulus derived from the light curve and $\mu_z$ is the distance modulus predicted given the redshift of the SN and Hubble constant $H_0$. 

It has been known for more than twenty years that rapidly declining (therefore low luminosity) SNe Ia are found more often in passive galaxies \citep{hamuy2000, ginolin2024a}, implying that the oldest SNe Ia progenitors have faint peak luminosities. A reasonable expectation could be that the relationship between the luminosities of SNe Ia and their light curve shape and color may depend upon progenitor age. Using SNe Ia observed as part of the Zwicky Transient Facility (ZTF), \citet{ginolin2024a} found stretch-magnitude and \citet{ginolin2024b} found color-magnitude relationships of SNe Ia change depending on the morphology of the host galaxy. Since morphological type is broadly correlated with galaxy age, this correlation could be related to progenitor age. 

As discussed above, HRs have shown a statistically significant dependence on H$\alpha$ surface brightness, which traces young stellar populations \citep{rigault2013, rigault2020}. However, it is worth noting that H$\alpha$ emission traces very recent star formation, on the order of 5-10 Myr \citep{calzetti2012}, and SNe Ia progenitors are thought to be at least 100 Myr old \citep[e.g.,][]{branch2017}. \citet{gallagher2008} and \citet{rose2019} both find evidence for a correlation between Hubble residuals and host-galaxy age inferred from fitting spectra of early-type host galaxies. Their results may support the idea that progenitor age is responsible for the host-galaxy mass dependence in HRs for SNe found in early-type host galaxies. However, \citet{dixon2022} find that the HRs of SNe Ia correlate with emission line strength but \textit{not} the stellar age inferred from fitting their host-galaxy spectra.  

One simple and common parameterization of the dependence of SNe Ia Hubble residuals on their host galaxy properties is a step function, which we will use in our paper. Several analyses \citep[e.g., ][]{sullivan2010, childress2013, johansson2013, rose2019} find that the relation between HRs and galaxy properties is nonlinear and prefer a step model or another model with a sharp transition. 

A step function of Hubble residuals as a function of stellar mass could be expected if the origin of the host-mass dependence is the age of the progenitor \citep{childress2014}. \citet{mannucci2006} suggest that there may be two populations of SNe Ia, specifically prompt and delayed: that is, SNe Ia that occur soon after a star becomes a white dwarf, and those that occur much later. Later efforts, including \citet{childress2013}, \citet{rigault2020}, and \citet{briday2022}, may support a bimodal distribution of progenitor ages. \citet{childress2014} find that the mean ages of SN Ia progenitors could be expected to exhibit a sharp transition between SNe Ia in lower-mass galaxies, which have higher specific star formation rates (SFRs), and higher-mass galaxies, which have lower specific SFR.
However, many galaxy properties, including stellar mass, SFR, metallicity, and dust content, are mutually correlated, which makes identifying a physical origin for the host dependence challenging. 
%model the star formation histories of galaxies and convolve them with theoretical delay-time distributions (DTDs, models of SNe Ia rate versus progenitor age) to demonstrate that the bimodal distribution we observe could arise  from different rates of star formation in low- versus high-mass galaxies, yielding the apparent dependence on host galaxy stellar mass as well as SFR.

The properties of interstellar dust are known to vary across astrophysical environments, and a reasonable question is whether the host-galaxy dependence could be connected to differences in the properties of dust. Interpretation of observations of SNe Ia and their host galaxies, however, has not yet yielded a consensus. \citet{brout2021} and \citet{meldorf2022} both find that they do not need to invoke intrinsic variation in SN Ia luminosity after light curve correction to explain the host-galaxy dependence; dust alone can account for all the differences in SNe Ia. On the other hand, %\citet{childress2013} and 
\citet{duarte2023} find that even after accounting for variation in the dust attenuation law, a step in Hubble residuals with host-galaxy stellar mass persists. Since dust extinction is much smaller at near-infrared (NIR) wavelengths than at optical wavelengths, SNe Ia distances  measured from NIR observations may be able to identify whether dust variation can explain the host-galaxy mass step. However, the observational evidence is equivocal. \citet{uddin2020} and \citet{ponder2021} find using NIR observations that the host mass dependence is still present, implying that variation in dust is not a strong contributor to the step. However, \citet{johansson2021} do not identify a dependence of Hubble residuals on host-galaxy stellar mass from NIR observations. 

Taken in aggregate, these apparently conflicting measurements present a puzzle for those looking to understand and best correct for the dependence of SNe Ia luminosities on their host-galaxy environment. In this paper, we approach the problem from a local perspective, using measurements of the environment around the supernova, comparable to \citet{rigault2013} (and the updated results in \citet{rigault2020}), \citet{jones2018}, and \citet{roman2018}. We use a sample of SNe Ia in host galaxies with star-forming morphology, similar to the sample-selection criteria used by the SH0ES team \citep{riess2022}. Our sample of SNe is primarily drawn from those with light curves published by the Lick Observatory Supernova Search (LOSS), Harvard-Smithsonian Center for Astrophysics (CfA), and Carnegie Supernova Project (CSP) surveys. We observed the host galaxies of these SNe Ia, which we call the North+Lone Star Survey. We measure the surface brightness of H$\alpha$ within a radius of 1 kpc of the SN explosion site. The HRs are from the Pantheon+ sample. We repeat the analysis using distances from  \citet{betoule2014} and \citet{hicken2014}. The paper is organized as follows. Data collection and reduction are discussed in section \ref{sec:data}. Section \ref{sec:analysis} describes our analysis of the data. In Section \ref{sec:results}, we discuss our results. Finally, Section \ref{sec:discussion} presents our conclusions. This is the first of a series of papers where we will present analysis of our IFU observations of nearby SNe Ia in the Hubble flow.

\section{Data} \label{sec:data}

\subsection{Data Collection}
We acquired spectroscopy of host-galaxy targets over a total of 48 nights from April 21, 2012 to March 11, 2024 using the George and Cynthia Mitchell Spectrograph (GCMS; \citealt{VIRUSP}) (previously known as Visible Integral-field Replicable Unit Spectrograph--Prototype, or VIRUS-P) mounted on the 2.7-m Harlan J. Smith telescope at McDonald Observatory. GCMS is an integral field unit (IFU) instrument comprised of 246 fibers, and each fiber is approximately 4.16 arcseconds in diameter. The field covers a 1.7$'$ x 1.7$'$ area. 

Each host galaxy was observed at three dither positions. The second dither was offset by $\Delta\alpha$ = -3$''$ and $\Delta\delta$ = -2$''$ from the first pointing, and the third dither was offset by +1.5$''$ and -4$''$, respectively, from the first pointing. Since GCMS has a 1/3 filling factor, a set of three dithers is required to cover entirely the footprint on the sky \citep{VIRUSP}. We refer to each exposure as a frame. A set of frames consists of two exposures taken at each of three dither position. 

We observed the host galaxies of 146 Type Ia SN, and 111 of these are in the Pantheon+ sample. The initial sample of 133 SNe was selected by visual inspection to include only those with a star-forming morphology to allow for gas-phase metallicity measurements and as a match to the criterion used by the SH0ES analysis \citep{riess2022}. We added thirteen early-type galaxies later for reasons that we describe in Section \ref{sec:discussion}. 

In this paper, which is the first of a series of papers, we present measurements from our data acquired in our ``red'' setup with a wavelength range of 4700-6590 \AA. The wavelength coverage allows for measurements of both H$\alpha$ and H$\beta$ as well as [N II]$\lambda 6583$, [O III]$\lambda 5007$, and the [S II] doublet $\lambda\lambda 6716, 6731$. In this paper, we measure the H$\alpha$ surface brightness. While we also observed these galaxies in the blue setup (3700-5950 \AA), here we analyze the red setup. 

\subsection{Initial Reduction}

All data were processed through {\tt Vaccine}, which is the data analysis pipeline for GCMS \citep{Vaccine}. {\tt Vaccine} performs bias correction, flat fielding, wavelength calibration, background subtraction, and cosmic ray rejection.

We completed the next data-processing steps using software routines from the VIRUS-P Exploration of Nearby GAlaxies (VENGA) program \citep{venga2}. Relative flux calibration was performed using the set of twelve standard stars listed in Table \ref{tab:standard}. We used an {\tt IDL} routine created by the VENGA collaboration \citep{venga2} to measure the response function from our observations of standard stars and to calibrate the spectra \citep{standards}.

\begin{table}[!htbp]
\centering
\caption{Standard stars used for calibration with corresponding setup. B corresponds to the blue setup, and R corresponds to the red setup.}
\begin{tabular}{| c | c  c|}
\hline
Star Name & \multicolumn{2}{|c|}{Setups Used} \\
\hline
BD +17\,4708 &   & R \\
BD +26\,2606 &   & R \\
BD +28\,4211 & B & R \\
BD +33\,2642 & B &   \\
Feige 110    & B &   \\
Feige 34     & B & R \\
Feige 56     & B & R \\
G191-B2B     & B & R \\
HD19445      &   & R \\
HD84937      &   & R \\
Hilt600      &   & R \\
HR1544       &   & R \\
HZ44         & B &   \\
\hline
\end{tabular}
\label{tab:standard}
\end{table}

We compared the response functions from individual standard star observations to evaluate the stability of the instrument and to identify potential problems. We looked at the flux calibration values at 6700\AA, which corresponds to the wavelength of H$\alpha$ at $z\approx0.02$. We normalized the curves by the median value of the flux in each curve and found that over the course of all of our runs, the values at 6700\AA\ have a standard deviation of 3.8\% erg$\times$cm$^{-2}\times$ ADU$^{-1}$ with a maximum fractional difference of 13\%. Within a single run, we find that the unnormalized values at the same location have a standard deviation of 1.5\% with a maximum fractional difference of 15\%. 

When compared to the response functions measured using other standard stars, the response functions for nights in the red setup for the standard BD +174708 were consistently smaller (at most $\sim$20\% at wavelengths below 5000\,\AA, where normalized values are $\approx1$). We therefore used alternative standard stars acquired during the same observing run to calibrate the data, since the response function is consistent within each run. A single night calibrated with BD +17\,4708 (2014-09-03) lacked an alternative standard star observation in the red setup, so we used a response function measured during an observing run taken two months previously.

\subsection{Astrometry and Flux Calibration}
We next applied a VENGA routine to perform astrometry and flux calibration simultaneously \citep{venga2}. The routine first uses the extracted spectrum to compute synthetic flux densities in the {\it g} or {\it r} band for each fiber using the filter's transmission function \citep{venga2}. A calibrated image from e.g., the Sloan Digital Sky Survey (SDSS), is next used to calculate the expected flux density for each fiber in each filter. The calibrated reference image is convolved with a Gaussian in order to match the seeing during the observation, and then the expected flux density for each fiber is computed from the convolved image \citep{venga2}. 

The routine simultaneously varies the astrometric solution and flux calibration of the IFU data cube. At each step, it computes the $\chi^2$ goodness-of-fit agreement between the shifted and rescaled datacube and the convolved, broadband {\it g} or {\it r} band image using the synthetic flux densities for each fiber and the expected flux densities from the calibrated image, as discussed in the previous paragraph.
The program then evaluates the $\chi^2$ statistic across a grid of parameter values iteratively over three searches, with each iteration starting from the optimal point found by the previous step. In the last step, the routine searches a higher resolution grid to determine with greater precision the best-fitting values of the parameters. In addition to fitting for the position of the fiber array on the sky, it performs a linear fit of the form $y=mx+b$, where $y$ is the flux density of the reference image and $x$ is the flux density from the observation. In this way, the program is able to perform an absolute flux calibration through comparison to the calibrated survey images as well as obtain precise astrometry required when combining the data acquired at multiple dither positions. 

Reference images used in this process were taken from from Pan-STARRS 1 (PS1) \citep{PS1-1, PS1-5, PS1-6, https://doi.org/10.17909/s0zg-jx37}, Sloan Digital Sky Survey (SDSS) \citep{SDSS1, SDSS2, SDSS_photo}, and the Zwicky Transient Facility (ZTF) \citep{ZTF}. The host-galaxy spectra of the following SNe had a signal-to-noise ratio (SNR) that was not sufficient to apply the astrometry and photometric calibration routine: SN1997dg, SN2005hn, SN2006ax, SN2006er, SN2006on, SN2007jg, SN2007lt, SN2007pf, SN2008fr, SN2008hm, SN2016blg, and Gaia16acv. Two host galaxies (SN2001V, SN2007bd) did not have reference images. Two SNe (SN2000dl, SN2013be) were at too great a redshift for GCMS to capture H$\alpha$. One object (SN2011df) was incorrectly aligned and did not contain the local environment of the SN. Observations of the thirteen remaining host galaxies failed to yield a consistent astrometric calibration: SN2000ce, SN2000fa, SN2001gb, SN2002he, SN2003he, SN2004as, SN2005eq, SN2005hf, SN2006cj, SN2006en, SN2011bc, SN2011ho, SN2013bs, and SN2013bt. 

\subsection{Data Cubes and Spectral Fitting}
Using an additional VENGA routine, we next combined  data frames from each observation to create a data cube \citep{venga2}.   

\begin{figure*}
    \centering
    \includegraphics[width=\textwidth]{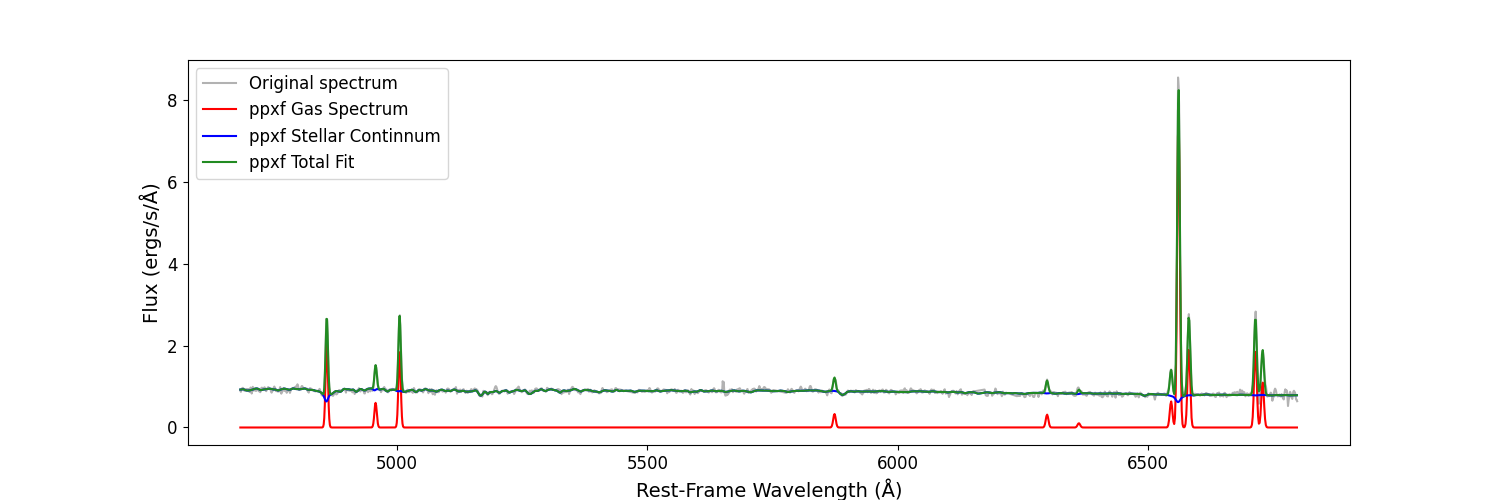}
    \includegraphics[width=\textwidth]{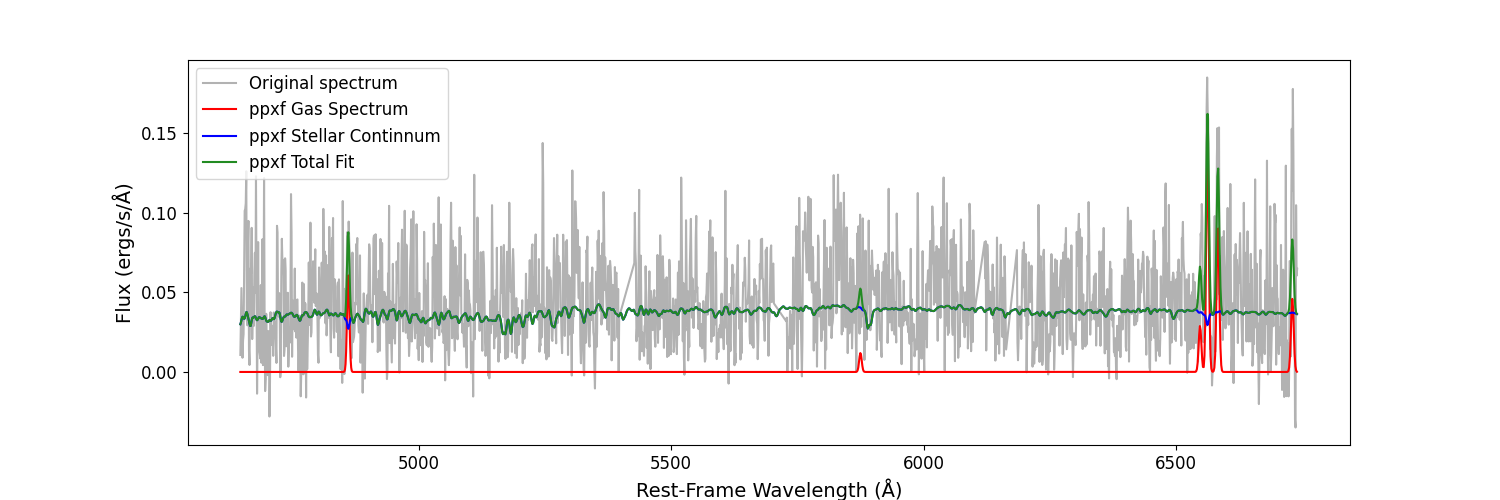}
    \caption{Example GCMS spectra within 1\,kpc of the SN explosion site with low and high S\slash N ratio. The upper panel shows the high S\slash N spectrum of the local environment of SN2004bg. The lower panel plots the spectrum for SN1990O, which has a lower S\slash N. The gray lines correspond to the raw data, while green lines denote the model. The blue line shows the contribution of the continuum, while the red shows the nebular emission.}
    \label{fig:spectra}
\end{figure*}

\begin{figure*}
    \centering
    \includegraphics[width=0.24\textwidth]{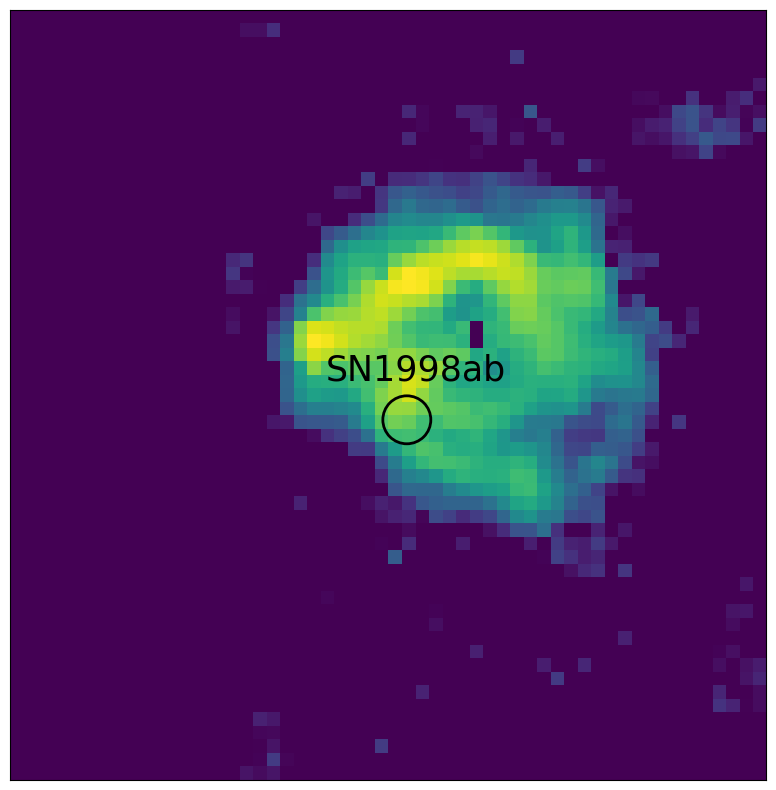}
    \includegraphics[width=0.24\textwidth]{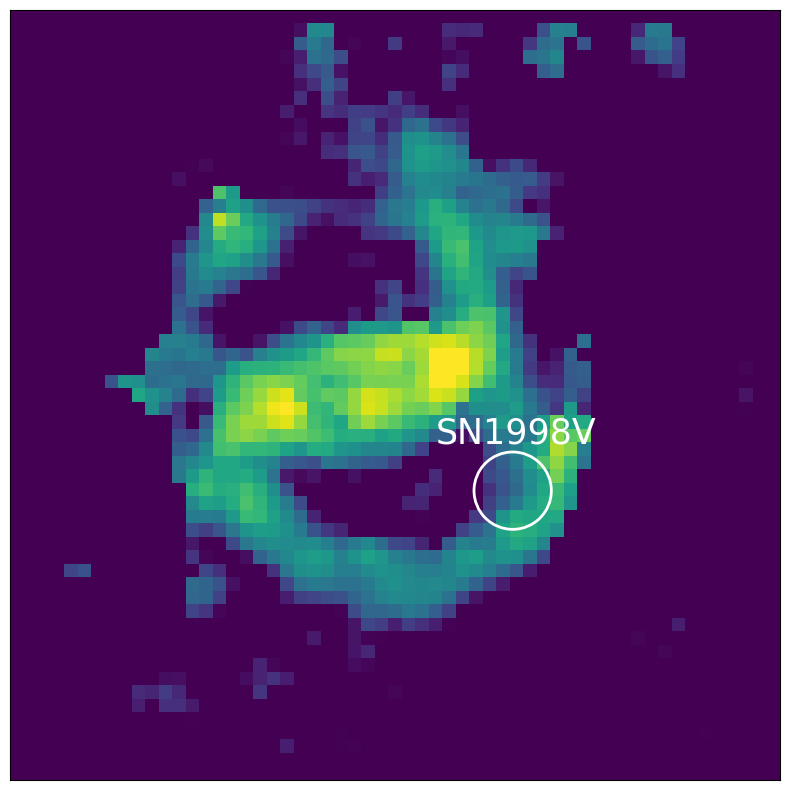}
    \includegraphics[width=0.24\textwidth]{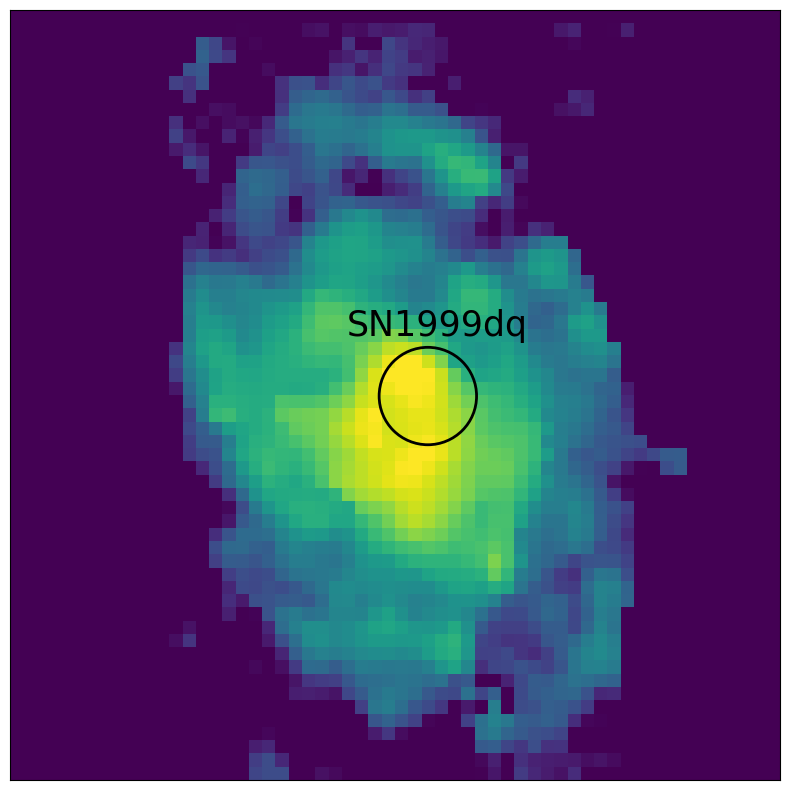}
    \includegraphics[width=0.24\textwidth]{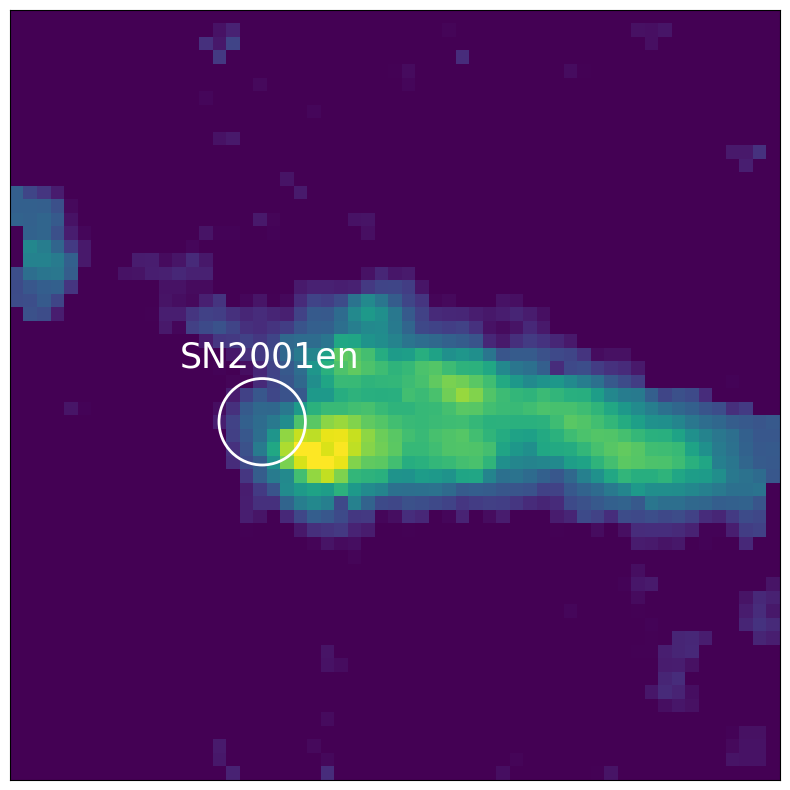}
    \includegraphics[width=0.24\textwidth]{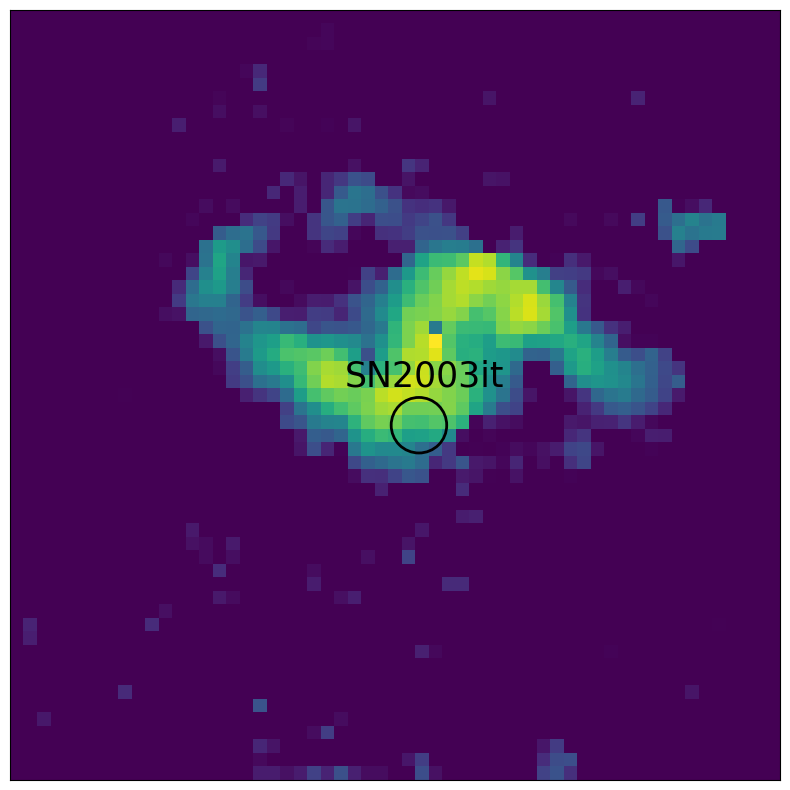}
    \includegraphics[width=0.24\textwidth]{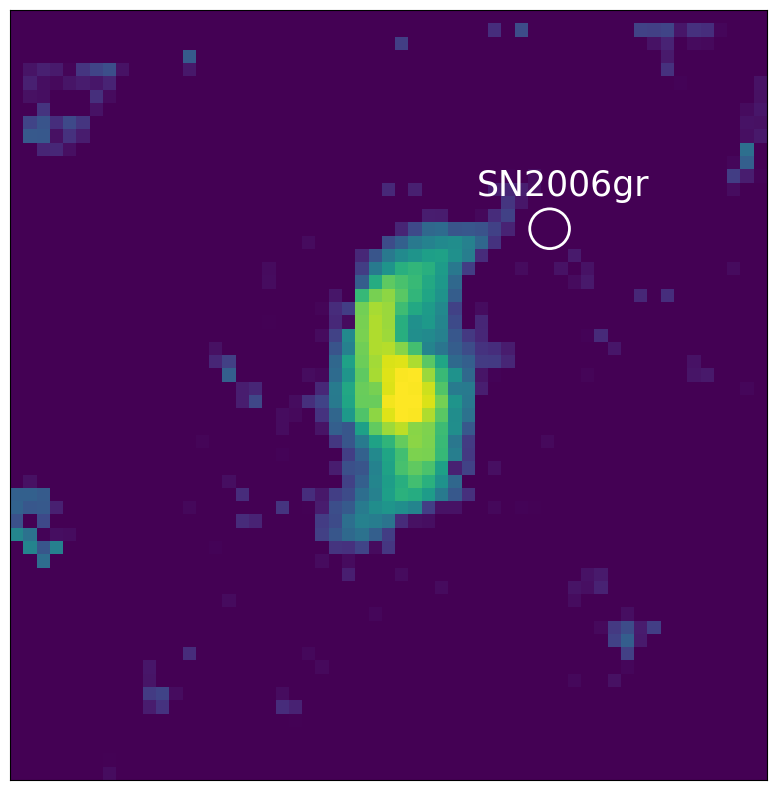}
    \includegraphics[width=0.24\textwidth]{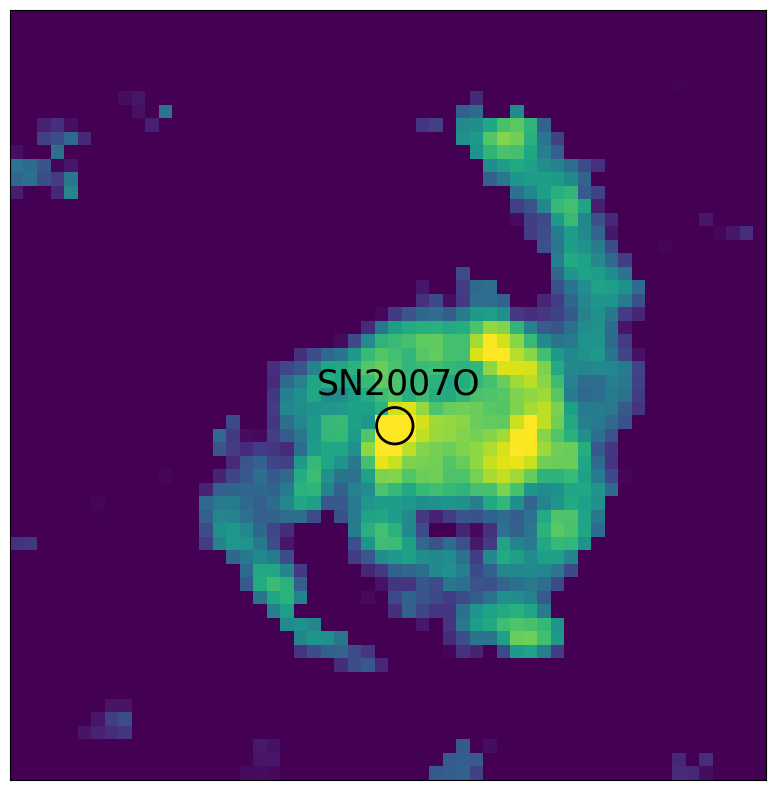}
    \includegraphics[width=0.24\textwidth]{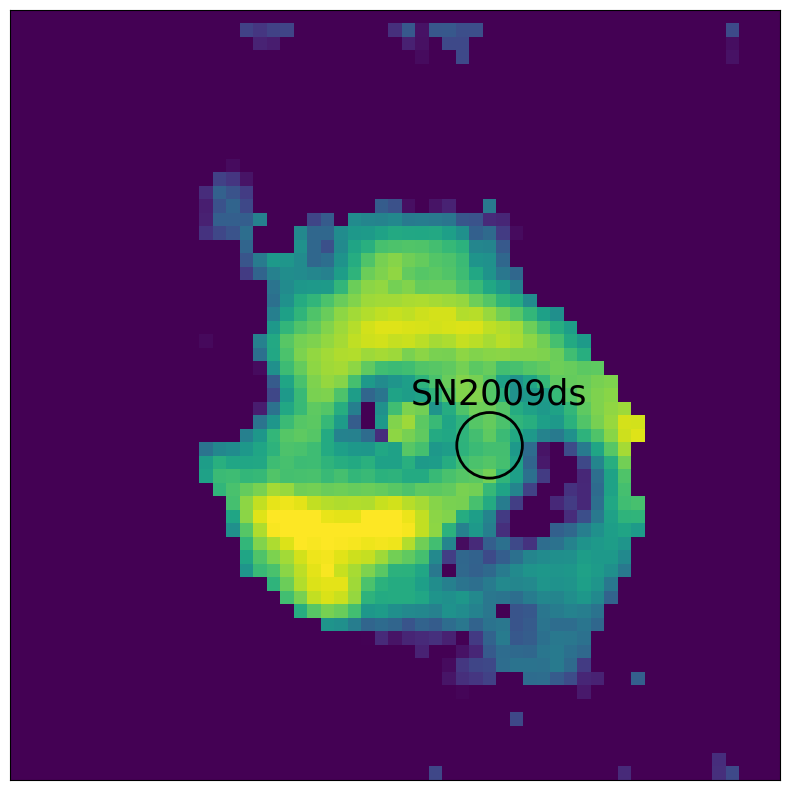}
    \caption{Example H$\alpha$ maps created using pyPipe3D \citep{pyPipe3D}. %for visualization purposes. 
    The location of the SN in each image is marked with a white or black circle and labeled with the name of the SN. The radius of the circle corresponds to the extraction radius that we use when extracting the spectrum.}
    \label{fig:maps}
\end{figure*}

Data cubes were analyzed using the Penalized PiXel-Fitting (\texttt{pPXF}) method  \citep{cappellari2023, cappellari2017, cappellari2004}, a data analysis package that fits a galaxy spectrum with a linear combination of simple stellar population models in combination with nebular emission, where the kinematics and age distributions of the stellar population as well as nebular kinematics and line strengths are free parameters. We retrieved the coordinates and redshift of each SN from the NASA Extragalactic Database (NED). We extracted the spectrum of the host galaxy environment within a circle with a 1\,kpc radius centered at the SN position. We computed the radius corresponding to 1\,kpc in arcseconds using the angular diameter distance at the redshift of the host galaxy. We then used the redshift of the host galaxy to compute the wavelengths of the spectrum in its rest frame and performed fitting using \texttt{pPXF}. We calculated uncertainties on the line strengths using bootstrapping with replacement and repeated \texttt{pPXF} fitting for 100 bootstrapped samples. Each fit was carefully inspected visually. Examples of our spectra are shown in Figure \ref{fig:spectra}, and H$\alpha$ maps are displayed in Figure \ref{fig:maps}.

Six SNe required a recalculation of redshift to match the wavelength of the strong nebular emission lines at the location of the SN: SN2001dl, SN2002de, SN2002jy, SN2003U, 2006mp, and SN2010ko. The rotation of the galaxy means that the recessional velocity at the location of the SN may differ from that of the galaxy nucleus.  

SN2010ko is adjacent to a spiral arm of NGC 1954 at $z=0.0105$. We find that its position is coincident with a compact background galaxy with redshift 0.0485. The spectrum of the SN itself shows that it is at a redshift of $z=0.0104$ \citep{cikota2019, parrent2010}. We excluded this object as its apparent local environment was contaminated by light from the background galaxy.

We calculated the H$\alpha$ surface brightness using the  line strength $F_{{\rm H}_{\alpha}}$ from \texttt{pPXF} together with the luminosity distance $D_L$, and $r$, the radius of extraction aperture in kpc,

\begin{equation}
    {\rm SB}= \frac{F_{{\rm H}_{\alpha}} \times 4 \pi D_L^2}{\pi r^2}.
    \label{eqn:SB}
\end{equation}

We calculated the angular radius of our aperture $r_{\theta}$ using the angular diameter distance at the host-galaxy redshift and the physical radius of our circular aperture $r$. 

\subsection{Hubble Residuals}

The distances published as part of the Pantheon+ sample \citep{brout2022} are corrected for a mass step that depends on the SN color. The Pantheon+ team (Brout, personal communication) provided us with a set of distances computed using a simpler model for the mass step that lacks any dependence on SN color. In this file, the Pantheon+ team had corrected for their model mass step  $\gamma$ at $\log_{10}(M_{\odot})=10$ of $0.05425$, where the higher mass galaxies host brighter SNe. We added the step back in by subtracting $\gamma/2$ from the high host mass SNe and adding $\gamma/2$ to the low host mass SNe. We use the redshift from the Pantheon+ catalog. We calculated the distance modulus, $\mu_z$, given the redshift, assuming a flat $\Lambda$CDM cosmology with $\Omega_M=0.298$ \citep{scolnic2018} and a value for the Hubble constant of $H_0=73.04$ km s$^{-1}$ Mpc$^{-1}$ \citep{riess2022} with the astropy cosmology package \citep{astropy2013, astropy2018, astropy2022}. We list the redshifts, distances, and HRs we derive from this sample in Table \ref{tab:data1}. We include an additional uncertainty in redshift representing peculiar velocity of 350 km/s\citep{hicken2009}. Twenty-two SNe observed in our sample were not in the Pantheon+SH0ES sample, which left a final sample of \objects\ SNe. 

For the purpose of comparison, we also computed HRs from the \citet{betoule2014} sample and the \citet{hicken2009} sample. We calculated the expected distance to each SN given its host-galaxy redshift using the cosmological parameters in each respective paper. Forty-two of our SNe were also present in the \citet{betoule2014} sample, and forty-nine in the \citet{hicken2009} sample. We calculated the HRs for the purpose of comparison, but did not use these values in our main analysis. 

\subsection{Surface Brightness Comparison}

\begin{figure}
    \centering
    \includegraphics[width=0.48\textwidth]{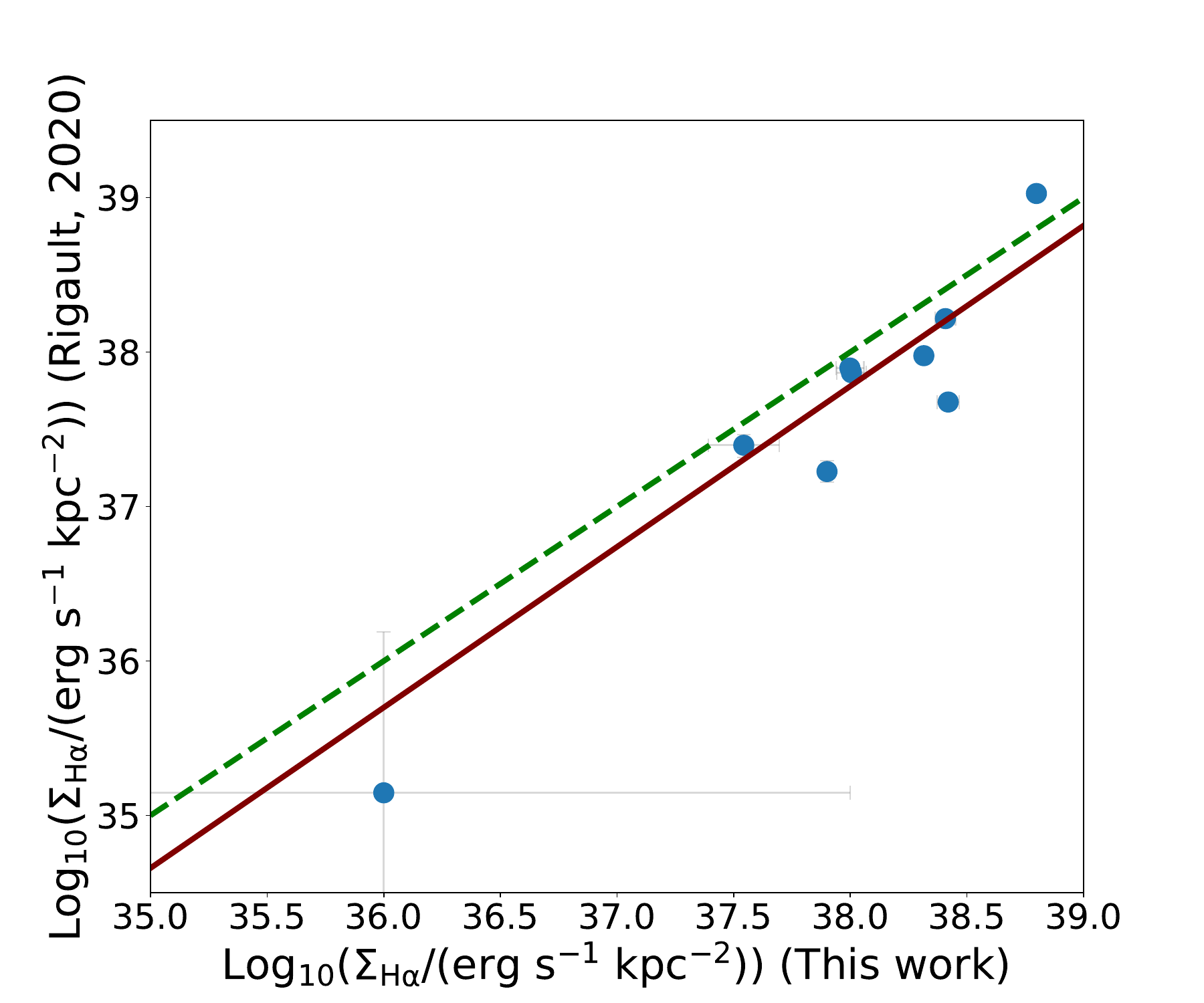}
    \caption{Comparison between our measurement of H$\alpha$ surface brightness and those published by  \citet{rigault2020} inside of the same 1\,kpc aperture. There are nine SN host galaxies that have both measurements. The solid maroon line plots the best fit, which has a slope of $1.04\pm0.24$ and a reduced $\chi^2$ of 3.73. The dashed green line shows the one-to-one line for reference. While the values are correlated and the slope is consistent with one, the measurements are not consistent within their uncertainties.}
    \label{fig:rigvus}
\end{figure}

We compared our measurements of the logarithm of H$\alpha$ surface brightness, \logHa, within a 1\,kpc radius aperture at the SN position both to those of \logHa\ in the same apertures from \citet{rigault2020} and to the SFR estimate in \citet{rigault2015} inferred from the far ultraviolet (FUV) imaging taken by the {\it Galaxy Evolution Explorer} ({\it GALEX}). Nine of the SN host galaxies in our sample are also in the \cite{rigault2020} sample. \cite{rigault2020} use the surface brightness of H$\alpha$ to estimate SFR per square kiloparsec, so we would expect our measurements to be in agreement. As shown in Figure \ref{fig:rigvus}, we find a slope of $1.04\pm0.22$ between our \logHa measurements and those from \citet{rigault2020}. While the slope is consistent with unity within less than 1$\sigma$, the differences between pairs of measurements of the same SN environment far exceed their combined uncertainties. We cannot account for the disagreement between the measurements of \logHa\ among the overlapping sample.

\begin{figure}
    \centering
    \includegraphics[width=0.5\textwidth]{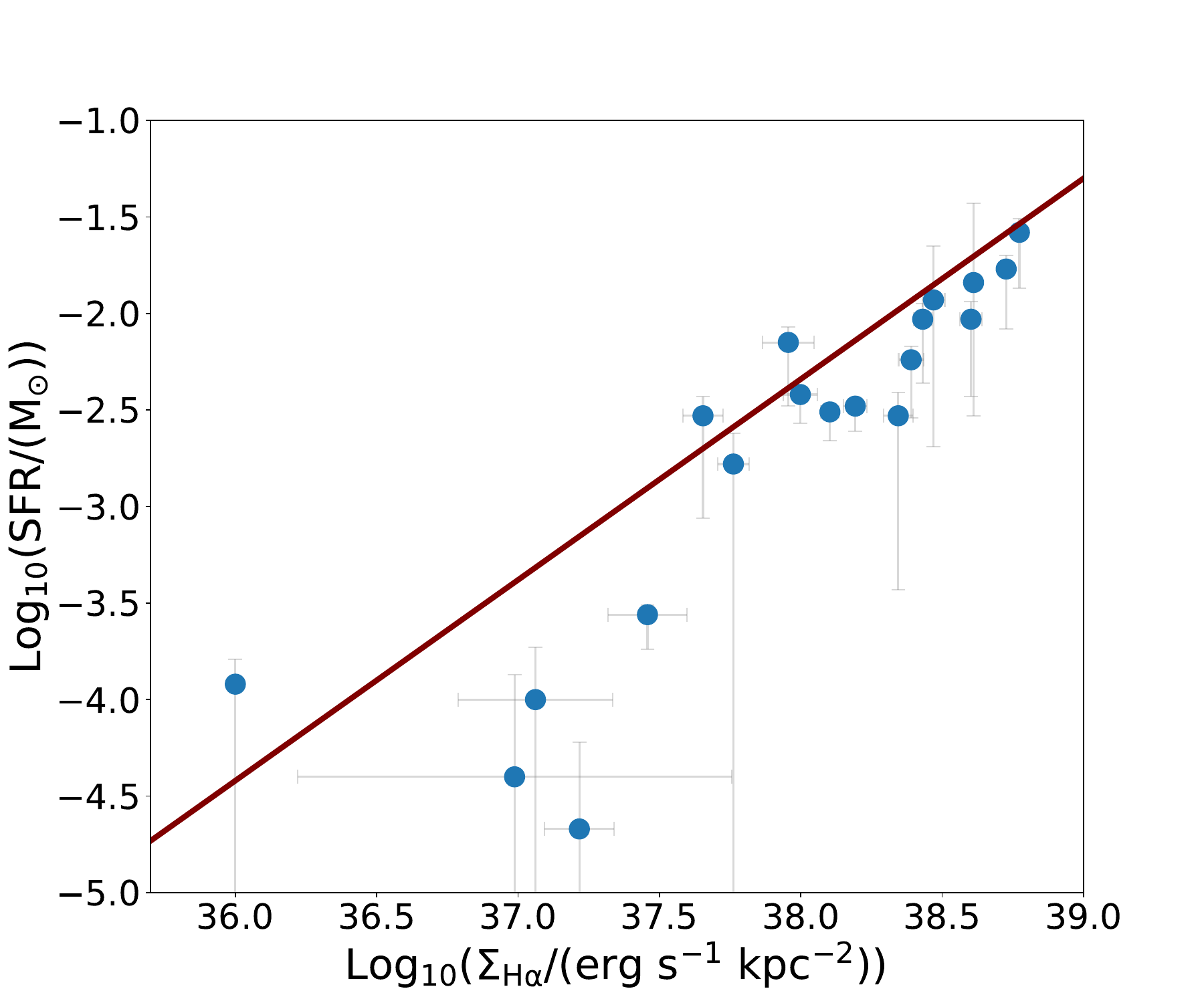}
    \caption{SFR as estimated by \citet{rigault2015} using ultraviolet flux from \textit{GALEX} with respect to our measurements of H$\alpha$ surface brightness. The maroon line shows the best fit, with a slope of $1.04\pm0.21$. The reduced $\chi^2$ value of the best-fit line is 0.55. The values are strongly correlated and the slope is consistent with the expected one-to-one comparison}
    \label{fig:GALEX}
\end{figure}

There are 19 host-galaxy environments in common between our sample and the \citet{rigault2015} \textit{GALEX} sample with estimates for SFR. We measure a slope of $1.04\pm0.21$ and plot the measurements in Figure \ref{fig:GALEX}. The slope is consistent with unity at the 1$\sigma$ level, as expected. 

\begin{figure}
    \centering
    \includegraphics[width=0.48\textwidth]{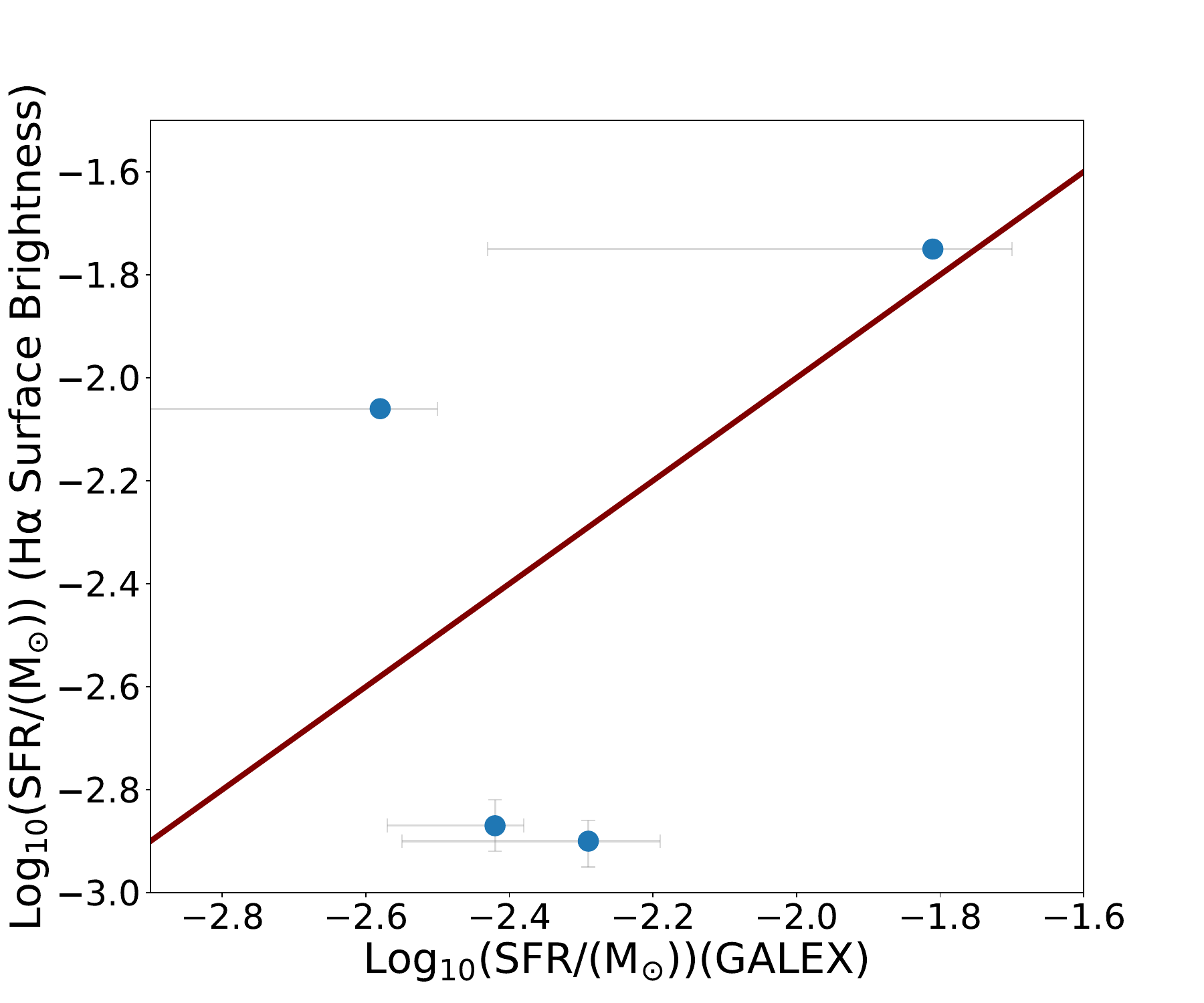}
    \caption{SFR estimated from \textit{GALEX} FUV imaging \citep{rigault2015} versus SFR inferred from H$\alpha$ \citep{rigault2020}. The red line is the one-to-one line for reference and does not represent a fit. The two measurements do not exhibit a strong correlation.}
    \label{fig:RigvGAL}
\end{figure}

While only four host galaxies are in both the \citet{rigault2015} and \citet{rigault2020} samples, we compared these SFRs estimated from H$\alpha$ and \textit{GALEX} FUV surface brightness, respectively. The comparison is shown in Figure \ref{fig:RigvGAL}, with the one-to-one line plotted in maroon for reference. While the correlation is not strong, the small sample size and large uncertainties for two of the measurements limits interpretation. 

\section{Analysis} \label{sec:analysis}

We compared the mean of Hubble residuals of SNe whose measurements of \logHa\ are below or above a split value. %in erg s$^{-1}$ kpc$^{-2}$. -- I don't think the units are necessary here
We adopt two locations for the step: the median of the \logHa\ measurements for our sample, and the median of the \citet{rigault2013} measurements, which they use in their analysis. %We split our data at these locations and then examined the properties of the SNe Ia on either side. 

We used the maximum likelihood method to constrain the size of the step in HRs, following \citet{jones2018} and \citet{rigault2020}. The method models the distribution of HRs as a piecewise function straddling the step location, or split point. Each side of the step is modeled as a separate Gaussian distribution with distinct mean and standard deviation. The probability that SN on each side of the step is computed using the uncertainty in \logHa. The log likelihood function, following \citet{jones2018} and \citet{rigault2020}, is 

\begin{multline}
    \sum_i \text{ln} \biggl( p_i\times\frac{1}{2\pi(\sigma_i^2+\sigma_a^2)}\text{exp}\left(\frac{-(\mu_a-x_i)^2}{(\sigma_i^2+\sigma_a^2)}\right)+\\
    (1-p_i)\times\frac{1}{2\pi(\sigma_i^2+\sigma_b^2)}\text{exp}\left(\frac{-(\mu_b-x_i)^2}{(\sigma_i^2+\sigma_b^2)}\right)\biggr)
\end{multline}

\noindent where $x_i$ is the HR of the $i$th SN, $\sigma_i$ is the uncertainty of the HR, $p_i$ is the probability that the local surface brightness lies below the split point, $\mu_a$ and $\sigma_a$ are the mean and standard deviation of the Gaussian distribution for SN HRs below the split point, and $\mu_b$ and $\sigma_b$ are the mean and standard deviation of the Gaussian distribution above the split point. We used Markov Chain Monte Carlo (MCMC) fitting to constrain the values of $\mu_a$, $\sigma_a$, $\mu_b$, and $\sigma_b$.

To compute the step, we subtract the mean of the HRs of the SNe whose host-galaxy measurements were smaller than the location of the split from the mean of those whose host-galaxy measurements were greater than the location of the split. We then calculated the mean of the residuals on each side of the step. We repeated this process $10^5$ times and calculated the median, standard deviation, and 16th and 84th percentiles of the resulting steps.  

In addition to applying the maximum likelihood method, we also calculated the weighted average of the measurements on either side of the HR step, using the inverse of the variance as the weight for each measurement. In contrast to the maximum likelihood calculation, this second calculation does not take into account uncertainty associated with the measurement of \logHa. Details of this calculation and results can be found in Appendix \ref{app1}. 

As a test of our implementation of the maximum likelihood method, we attempted to reproduce the results of \citet{rigault2020} using their measurements. We found a step with local sSFR  of $0.121 \pm 0.025$, while   \citet{rigault2020} reported $0.125 \pm 0.023$.

\begin{figure}
    \centering
    \includegraphics[width=0.5\textwidth]{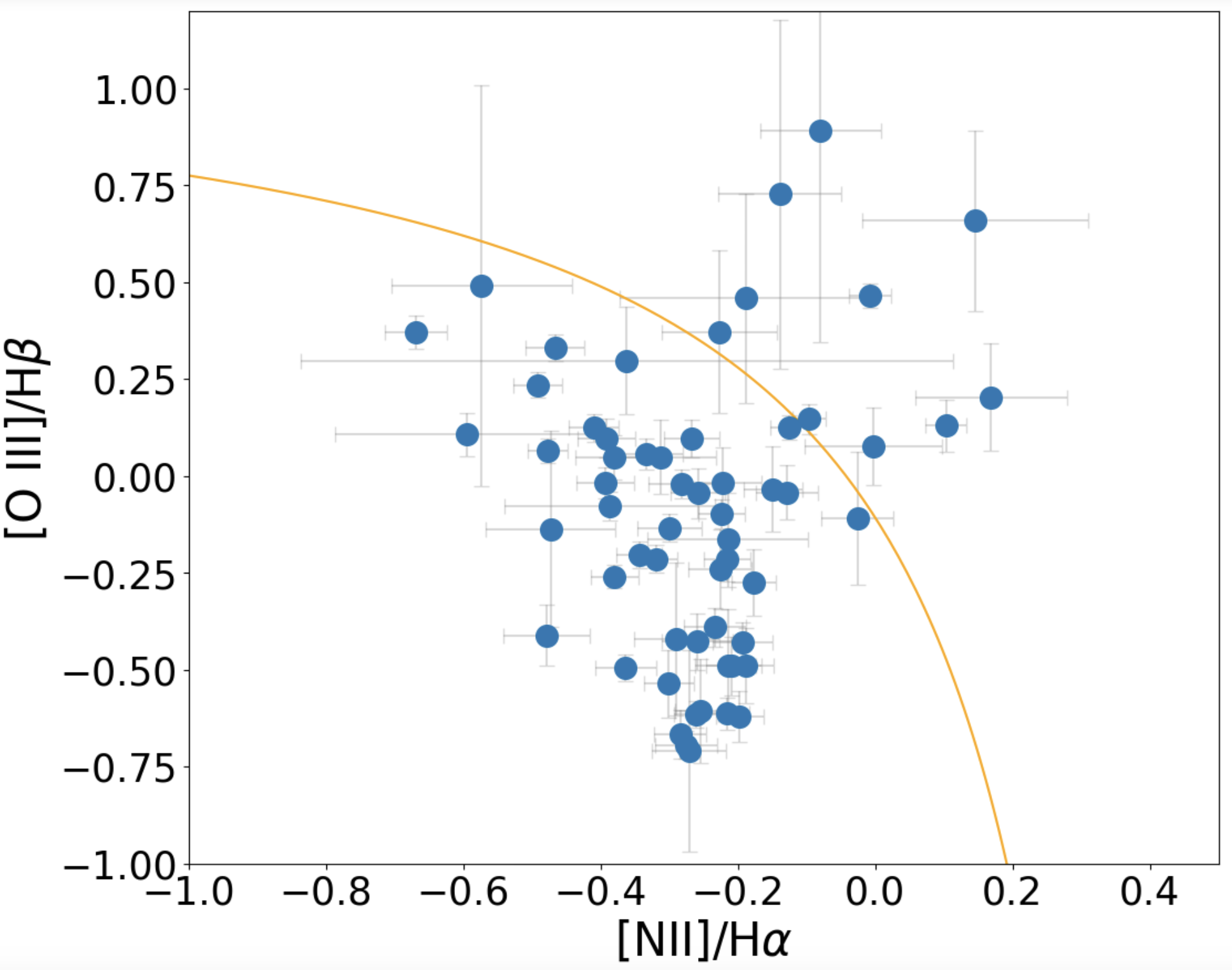}
    \caption{BPT diagram of our sample, showing the \citet{kewley2001} separation curve (orange). Local SN environments lying above this line are categorized as AGN, and SNe below it are star-forming or composite.}
    \label{fig:BPT_All}
\end{figure}

We next constructed a Baldwin-Phillips-Terlevich (BPT; \citealt{bpt1981}) diagram to assess whether line emission measured in the local apertures could be affected by contamination from Active Galactic Nuclei (AGN) or Low Ionization Nuclear Emission-line Regions (LINERs). We use the  \citet{kewley2001} partition between star-forming (below) and AGN (above) nebular emission line ratios. We first selected only local environments for which H$\alpha$, H$\beta$, [NII] $\lambda6583$, and [O III] $\lambda$5007 were detected at a signal to noise ratio (S/N) of greater than 1. Of the 55 SNe with S/N $>$ 1, 47 of our 82 SNe fall in the star-forming region (see Figure \ref{fig:BPT_All}). 

For our main sample, we used only SNe that passed SH0ES light curve cuts applied by \citet{riess2022}. SNe were removed from the SH0ES analysis based on color and stretch; specifically, only SNe with color $|c|<0.15$ and stretch $|x_{1}|<2$ were used in the cosmological analysis. The SNe we use are listed in Tables \ref{tab:data1} and \ref{tab:data2}. 

\section{Results} \label{sec:results}

\begin{figure*}[!h]
    \centering
    \includegraphics[width=0.45\textwidth]{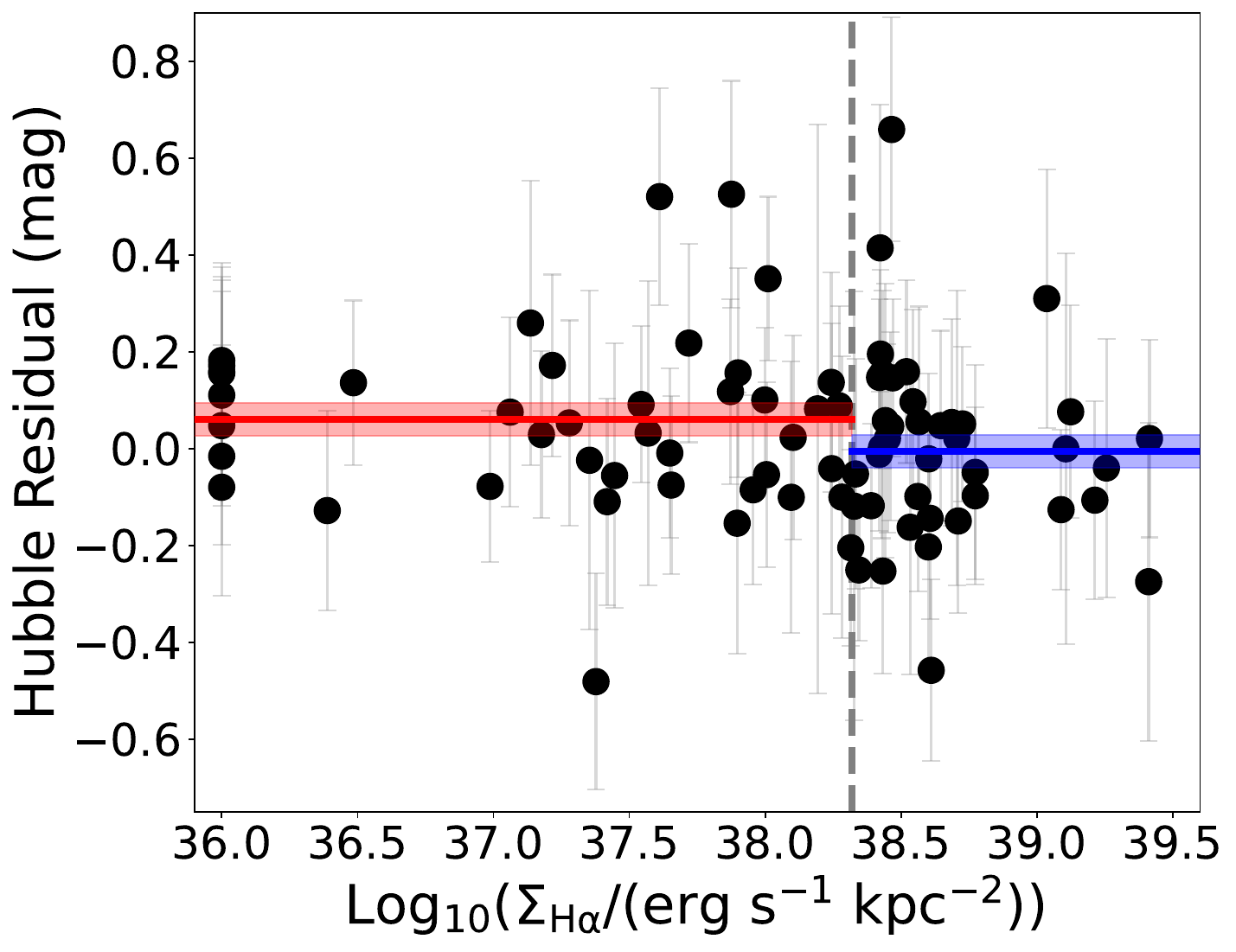}
    \includegraphics[width=0.45\textwidth]{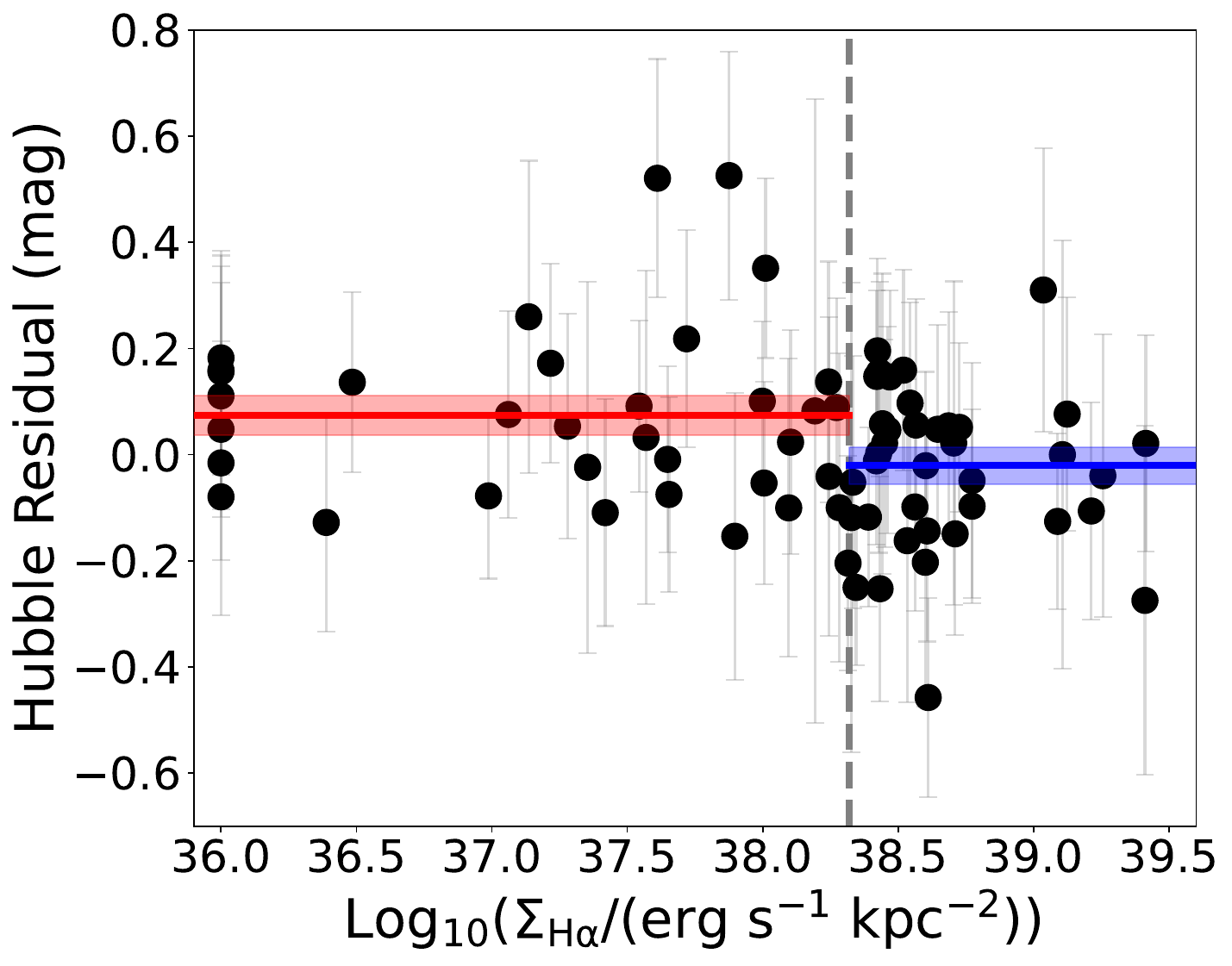}
    \caption{Hubble residuals plotted against \logHa. The gray dashed vertical line marks the location of the step in HRs, \logHa$=38.32$. This is the median value from \citet{rigault2013}. The solid red line shows the median HR of SNe with \logHa above the median; the solid blue line shows the median HR of SNe with \logHa below the median. The shaded regions show the 16th and 84th percentiles. The left figure shows all SNe; the right figure shows only the SNe that have passed light curve cuts.}
    \label{fig:Ha}
\end{figure*}

We measure a step in HRs of $-0.097 \pm 0.051$ mag with a significance of 1.9$\sigma$ when the split point is the median of the \citet{rigault2013} sample \logHa\ $=38.32$, where SNe in lower surface-brightness environments have more positive HRs, on average, than those in star-forming environments after correcting for the Tripp relation \citep{tripp1998}. Table \ref{tab:steps} lists our constraints for the multiple subsamples we consider. The significance of the step in \logHa\ never exceeds 2.5$\sigma$. Figures \ref{fig:Ha} shows the plot of \logHa\ for all SNe (left panel) and after light-curve cuts (right panel). We compare our results with those presented in \citet{rigault2020}, which uses the same \logHa measurements as \citet{rigault2013} but an updated version of SALT2 that has a greater similarity to the version used in the Pantheon+ analysis to calculate the HRs (see section 5.10 of \citet{rigault2020}). Both Pantheon+ and \citet{rigault2020} use the version of SALT2 presented in \citet{betoule2014}. However, the Pantheon+ team uses the Fitzpatrick extinction law for Milky Way dust extinction \citep{fitzpatrick1999} motivated by \citet{schlafly2011}, while the SALT2 base code used in \citet{rigault2020} uses the dust extinction from \citet{cardelli1989}. Our measurement is in 2.4$\sigma$ statistical tension with that of \cite{rigault2020} who found evidence for a step in the opposite direction of $0.045 \pm 0.029$ magnitudes. We also note that, in our sample, there are only six SNe with $\rm HR<-0.1$ and \logHa\ $<38.32$ (the lower left corner of both plots in Figure \ref{fig:Ha}), compared to \citet{rigault2013}, who find seventeen SNe Ia that meet these criteria. 

As shown in Table \ref{tab:steps}, applying the SH0ES light curve cuts based on color and stretch described in Sec.~\ref{sec:analysis} raises both the size and significance of the step. Applying BPT cuts also increases the size of the step, but decreases the significance, possibly a result of the smaller sample size. Using the median point of \citet{rigault2013}, \logHa$=38.32$, instead of the median of the data decreases the size of the step and its significance. 

\begin{table*}
    \centering
    \movetableright=-0.85in
    \begin{tabular}{|c|c|c|c|c|}
        \hline
        \multicolumn{5}{|c|}{All SNe} \\
        \hline
         Step Location (Log$_{10}$($\Sigma_{{\rm H}\alpha}$ [erg s$^{-1}$ kpc$^{-2}$])) & Additional Cuts & Sample Size & Step Size (mag) & Significance \\ \hline
         38.30 (This paper)         & ...         & 82 & -0.072 $\pm$ 0.048 & 1.5$\sigma$\\
         38.46 (This paper)         & BPT Non-AGN & 47 & -0.14  $\pm$ 0.070 & 2.1$\sigma$\\
         38.32 \citep{rigault2013}  & ...         & 82 & -0.069 $\pm$ 0.049 & 1.4$\sigma$\\
         38.32 \citep{rigault2013}  & BPT Non-AGN & 47 & -0.18  $\pm$ 0.098 & 1.8$\sigma$\\
        \hline
        \multicolumn{5}{|c|}{Light Curve Cuts Applied} \\
        \hline
         Step Location (Log$_{10}$($\Sigma_{{\rm H}\alpha}$ [erg s$^{-1}$ kpc$^{-2}$])) & Additional Cuts  & Sample Size & Step Size (mag) & Significance \\ \hline
         38.33 (This paper)         & ...         & 73 & -0.090 $\pm$ 0.051 & 1.8$\sigma$\\
         38.43 (This paper)         & BPT Non-AGN & 45 & -0.12  $\pm$ 0.069 & 1.8$\sigma$\\
         38.32 \citep{rigault2013}  & ...         & 73 & -0.097 $\pm$ 0.051 & 1.9$\sigma$\\
         38.32 \citep{rigault2013}  & BPT Non-AGN & 45 & -0.17  $\pm$ 0.093 & 1.8$\sigma$\\
        \hline
    \end{tabular}
    \caption{Measured \logHa\ step size and statistical significance for multiple samples and step locations. ``BPT Non-AGN" refers to the cuts using the \citet{kewley2001} line to exclude AGN. ``Light Curve Cuts" refers to the light curve cuts based on color and stretch detailed in Sec.~\ref{sec:analysis}. The first two lines in each section list our measurements when the step location is the median of the data. The following two lines list our measurements when using the step location used by \cite{rigault2013}.}
    \label{tab:steps}
\end{table*} 

\begin{figure*}
    \centering
    \includegraphics[width=0.48\textwidth]{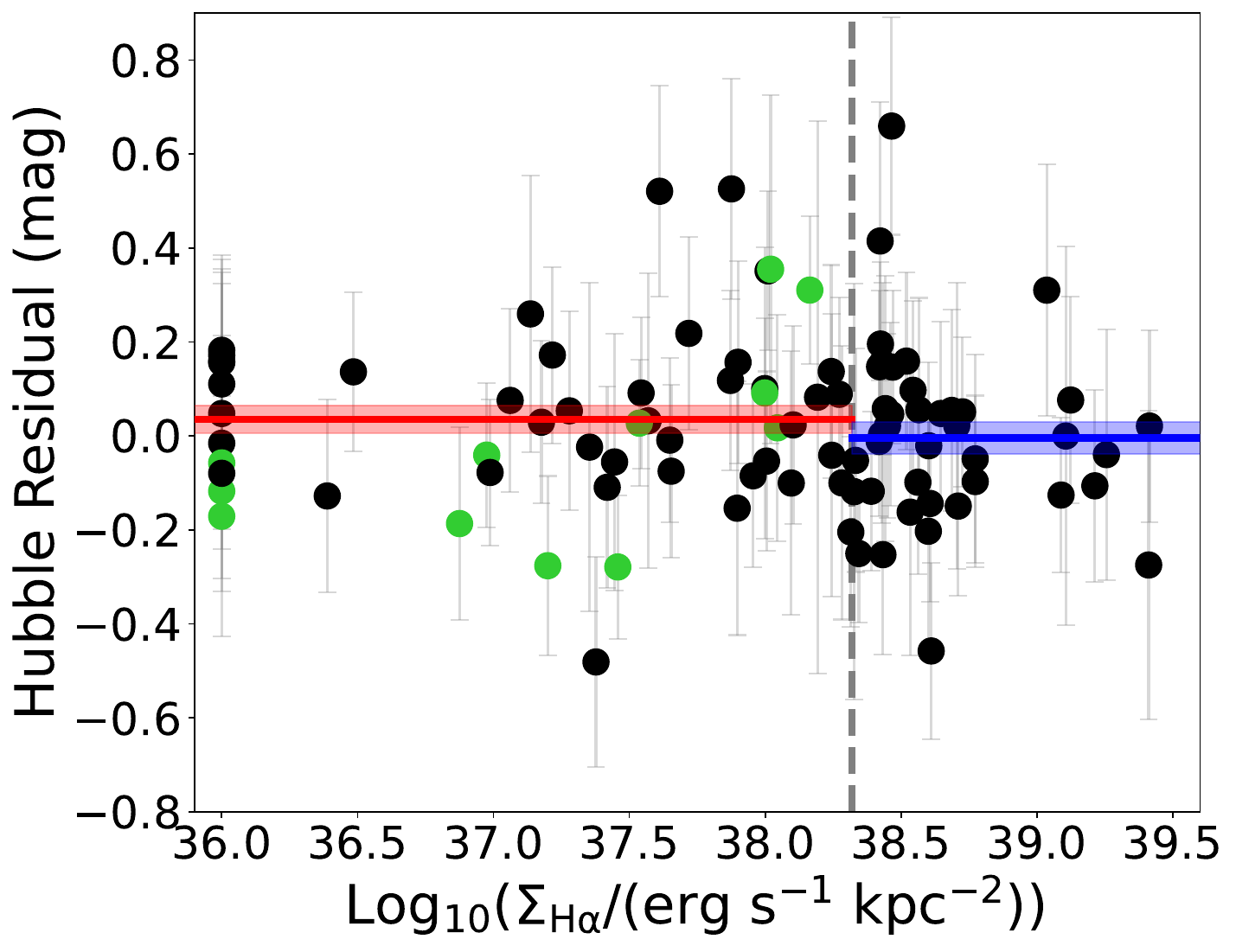}
    \includegraphics[width=0.48\textwidth]{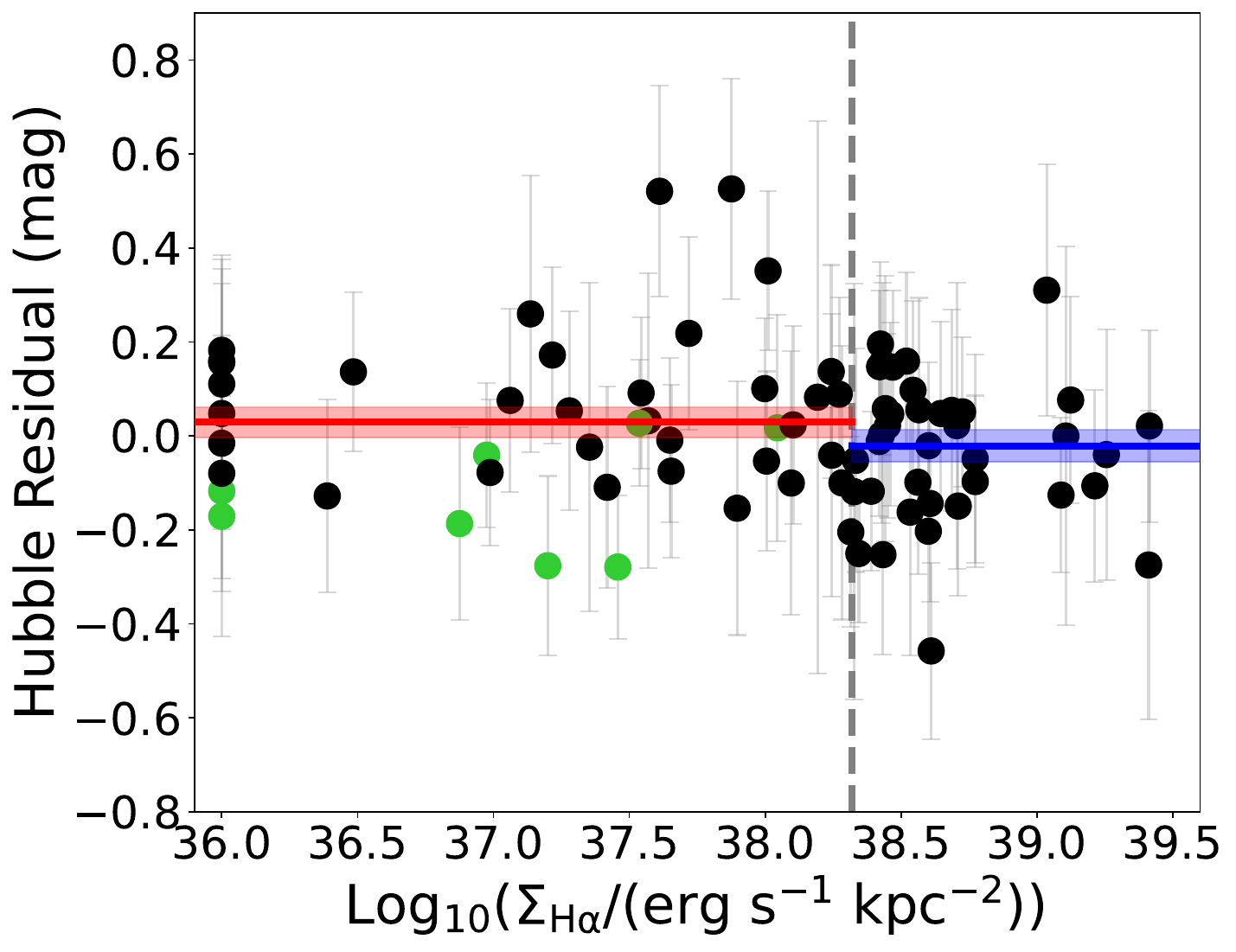}
    \caption{Similar to Figure \ref{fig:Ha}, but with early-type galaxies added. Early-type galaxies are shown in green; late-type galaxies are shown in black. It appears that the early-type hosts may fill in the relatively empty lower left quadrant of the plot.}
    \label{fig:ellipse}
\end{figure*}

We considered whether our selection of host galaxies with star-forming morphology could account for the tension between our results and those of \citet{rigault2020}. To investigate the possibility, we obtained IFU spectroscopy of the early-type host galaxies of 12 SNe, a mix of E and S0 type galaxies \citep{https://doi.org/10.26132/ned1}, and computed the step sizes after adding the measurements. Figure \ref{fig:ellipse} plots HRs against \logHa\, with host-galaxy morphology denoted by the points' colors. As expected, all early-type galaxies are on the ``passive'' side of the step. We find five additional SNe with HR$<-0.1$ and \logHa $<38.32$. The results of this analysis are listed in Table \ref{tab:ellipticals}. In the analysis with additional early-type galaxies we find a step of $-0.052 \pm 0.048$, in tension with \citet{rigault2020} at 1.7$\sigma$, a small decrease of 0.7$\sigma$. 

\begin{table*}[!htbp]
    \centering
    \movetableright=-0.85in
    \begin{tabular}{|c|c|c|c|c|}
        \hline
        \multicolumn{5}{|c|}{All SNe} \\
        \hline
         Step Location (Log$_{10}$($\Sigma_{H\alpha}$ [erg s$^{-1}$ kpc$^{-2}$])) & Cuts & Sample Size & Step Size (mag) & Significance \\ \hline
         38.13 (This work)         & ...         & 94 & -0.031 $\pm$ 0.045 & 0.70$\sigma$\\
         38.46 (This work)         & BPT Non-AGN & 48 & -0.15  $\pm$ 0.070 & 2.1$\sigma$\\
         38.32 \citep{rigault2013} & ...         & 94 & -0.044 $\pm$ 0.045 & 0.96$\sigma$\\
         38.32 \citep{rigault2013} & BPT Non-AGN & 48 & -0.19  $\pm$ 0.093 & 2.0$\sigma$\\
        \hline
        \multicolumn{5}{|c|}{Light Curve Cuts} \\
        \hline
         Step Location(Log$_{10}$($\Sigma_{H\alpha}$ [erg s$^{-1}$ kpc$^{-2}$])) & Cuts & Sample Size & Step Size (mag) & Significance \\ \hline
         38.27 (This work)         & ...         & 81 & -0.060 $\pm$ 0.047 & 1.3$\sigma$\\
         38.46 (This work)         & BPT Non-AGN & 45 & -0.12  $\pm$ 0.070 & 1.7$\sigma$\\
         38.32 \citep{rigault2013} & ...         & 81 & -0.052 $\pm$ 0.048 & 1.1$\sigma$\\
         38.32 \citep{rigault2013} & BPT Non-AGN & 45 & -0.17  $\pm$ 0.094 & 1.8$\sigma$\\
        \hline
    \end{tabular}
    \caption{Measurements of step size and significance for \logHa\ step with added early-type galaxies. ``BPT Non-AGN" refers to the cuts using the \citet{kewley2001} line to exclude AGN. \``Light Curve Cuts" refers to the light curve cuts based on color and stretch detailed in Sec.~\ref{sec:analysis}.}
    \label{tab:ellipticals}
\end{table*} 

We performed a test to examine the impact of the choice of the standard deviation of the peculiar velocity by calculating the steps again with a peculiar velocity standard deviation of 250 km/s. The step with respect to \logHa using the median from \citet{rigault2020} of \logHa$=38.32$ and after applying SH0ES light curve cuts based on color and stretch as describe in Sec.~\ref{sec:analysis} of $-0.097 \pm 0.051$ at 1.9$\sigma$ changes to $-0.096 \pm 0.051$ at 1.9$\sigma$. Our result with early-type galaxies added in with a peculiar velocity of 350 km/s is $-0.052 \pm 0.048$ at 1.1$\sigma$. With a peculiar velocity of 250 km/s, we find $-0.050 \pm 0.047$ at 1.1$\sigma$. We conclude that the choice of peculiar velocity does not have a significant impact on our results.

We also performed analyses using HRs instead from \citet{betoule2014} and \citet{hicken2009}. A total of 42 SNe were present in both our sample and in the \citet{betoule2014} sample. Using the Pantheon+ residuals, we find a step of $-0.069 \pm 0.070$ mag at 0.99$\sigma$ significance. Using the residuals calculated from \citet{betoule2014}, we find a step of $-0.053 \pm 0.058$ mag at 0.91$\sigma$ significance. The tension between these two values is 0.18$\sigma$. %, meaning the values are consistent with each other at the 1$\sigma$ level. 
Using the 49 SNe present in both our sample and in \citet{hicken2009}, the step with Pantheon+ residuals is $-0.056 \pm 0.063$ at a significance of 0.88$\sigma$ and the step with HRs from \citet{hicken2009} is $-0.030 \pm 0.063$ at a significance of 0.48$\sigma$. The tension between these two values is 0.29$\sigma$, meaning that the values are consistent at the 1$\sigma$ level. 

\begin{figure*}
    \centering
    \includegraphics[width=0.48\textwidth]{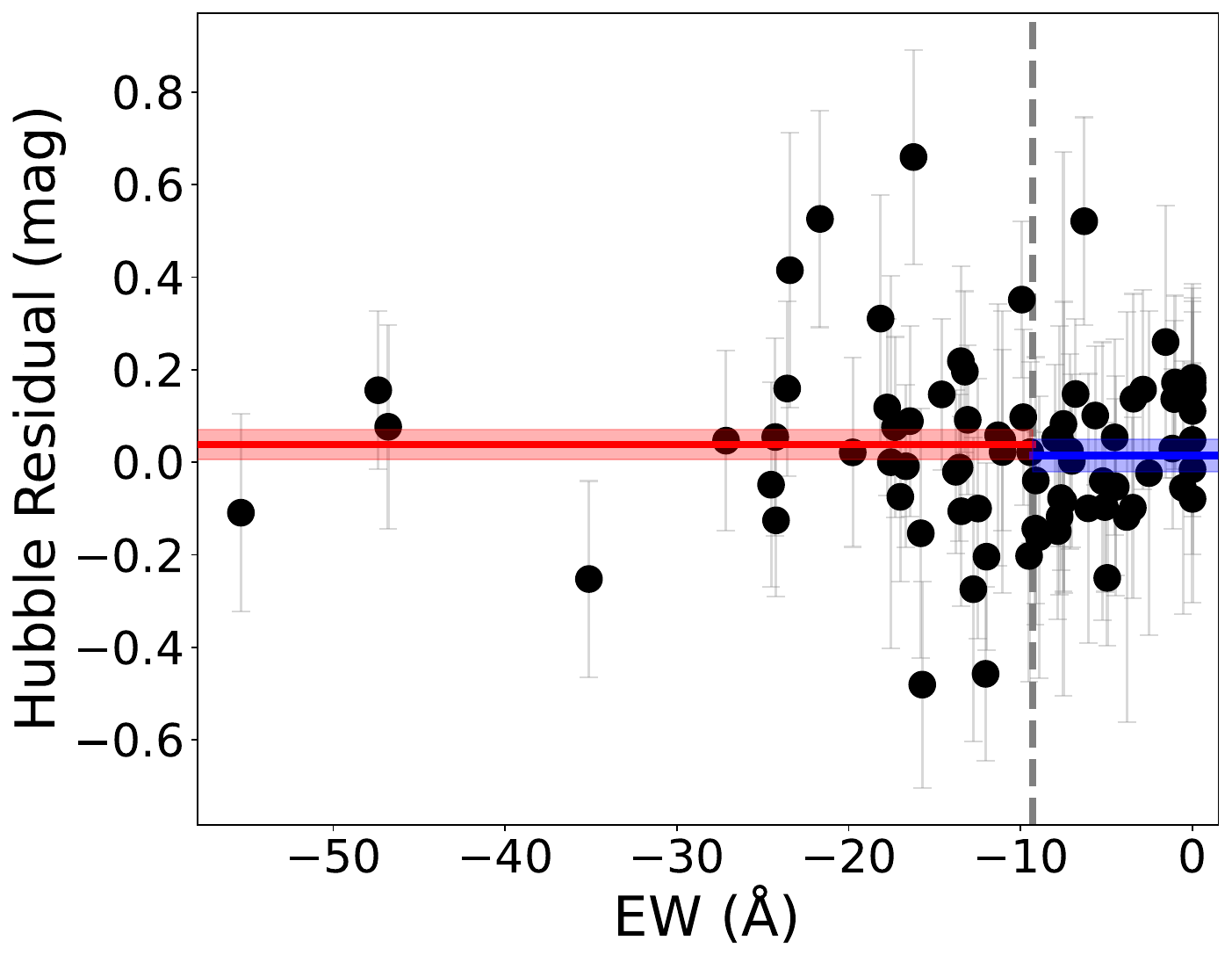}
    \includegraphics[width=0.48\textwidth]{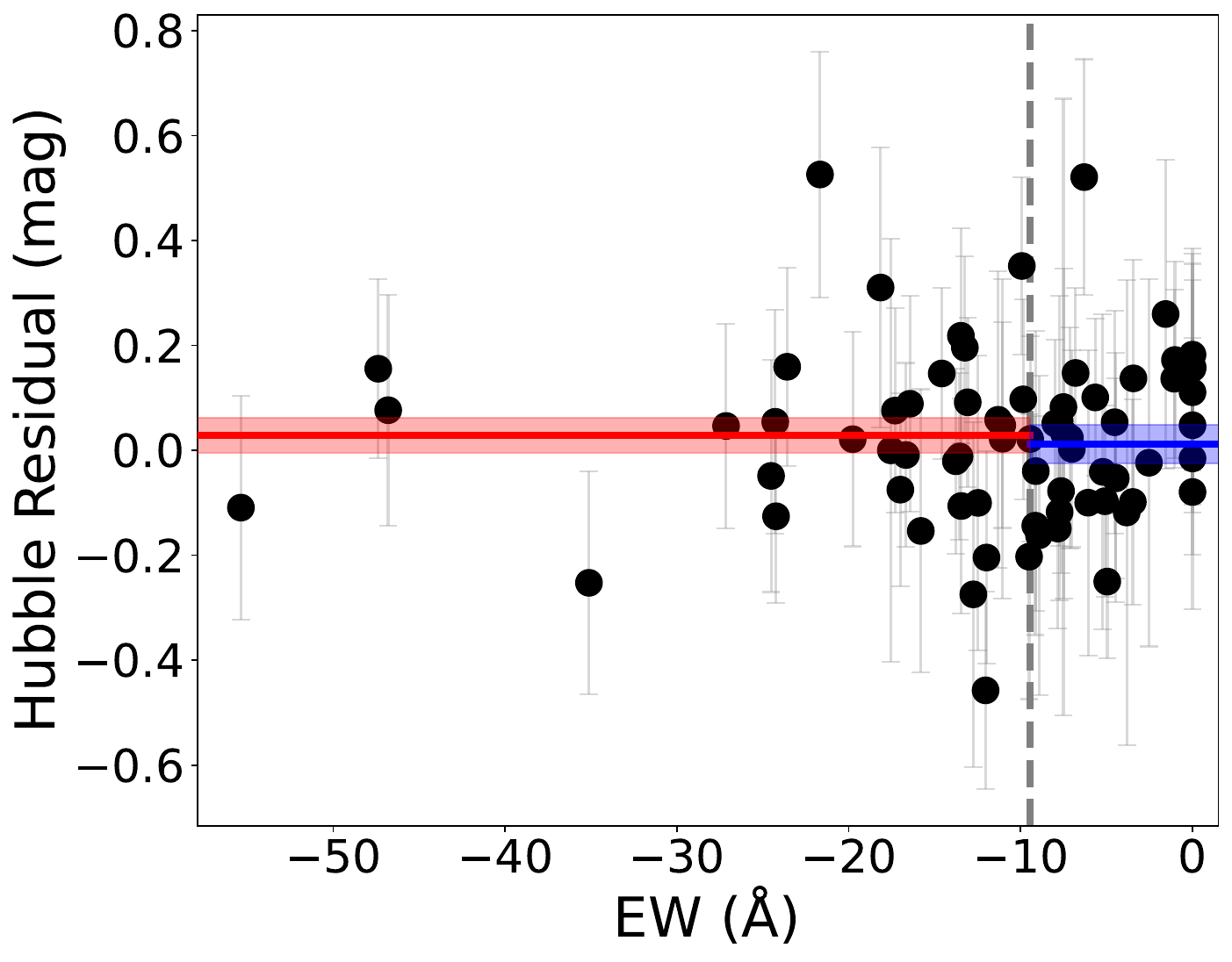}
    \caption{Hubble residuals with respect to the EW of H$\alpha$. The median is shown as a gray dashed line. The red solid line shows the median HR of SNe with EW above the median; the blue solid line shows the median HR of SNe with EW below the median. The shaded regions show the 16th and 84th percentiles. The left figure shows all SNe; the right figure shows only the SNe that have passed light curve cuts. The HR step is very small and not significant.}
    \label{fig:EW}
\end{figure*}

We also evaluated whether Hubble residuals are correlated with the equivalent width (EWs) of H$\alpha$ emission from the local environment.  For this analysis, we used the median of the EW measurements as the location of the step. The results are shown in Table \ref{tab:EW} and Figure \ref{fig:EW}. We find no significant evidence for a step in HRs with EWs, and no set of cuts resulting in a step with significance larger than 1$\sigma$. 

\begin{table*}[!htbp]
    \centering
    \movetableright=-0.9in
    \begin{tabular}{|c|c|c|c|c|}
        \hline
        \multicolumn{5}{|c|}{Host Galaxies Selected with Star-Forming Morphology} \\
        \hline
         Step Location (\AA) & Cuts & Sample Size & Step Size (mag) & Significance \\ \hline
         -9.30 & ...     & 82 & 0.019  $\pm$ 0.048 & 0.39$\sigma$ \\
         -9.45 & LC      & 73 & 0.015  $\pm$ 0.051 & 0.31$\sigma$ \\
         -12.0 & BPT     & 47 & 0.010  $\pm$ 0.070 & 0.14$\sigma$ \\
         -11.3 & BPT, LC & 45 & -0.032 $\pm$ 0.068 & 0.47$\sigma$ \\
        \hline
        \multicolumn{5}{|c|}{With Early-type Galaxies} \\
        \hline
         Step Location (\AA) & Cuts & Sample Size & Step Size (mag) & Significance \\ \hline
         -7.73 & ...     & 94 & 0.0052  $\pm$ 0.045 & 0.12$\sigma$ \\
         -7.99 & LC      & 81 & 0.027   $\pm$ 0.048 & 0.56$\sigma$ \\
         -11.6 & BPT     & 48 & -0.0085 $\pm$ 0.068 & 0.13$\sigma$ \\
         -11.3 & BPT, LC & 45 & -0.033  $\pm$ 0.067 & 0.49$\sigma$ \\
         \hline
    \end{tabular}
    \caption{Measurements of step size and significance for EW step with added early-type galaxies. ``BPT" refers to the cuts using the \citet{kewley2001} line to exclude AGN. ``LC" refers to light curve cuts based on color and stretch detailed in Sec.~\ref{sec:analysis}.}
    \label{tab:EW}
\end{table*} 

We also examined the stellar mass step as well as the $U-V$ color step using measurements from Table 7 in \citet{roman2018}. There were 27 SNe present in both samples. We set the split point at the median of the data for all parameters, as was done in \citet{roman2018}. The most comparable analysis performed in \citet{roman2018} is of SNe at $z<0.1$. The values from \citet{roman2018} are drawn from Table 7 of that paper, and tension was calculated using those values. In Table \ref{tab:roman}, we show the values from our analysis as well as the tension with the measurement from the \citet{roman2018} analysis of all Ia SNe at $z<0.1$. We note that \citet{roman2018} uses a 3\,kpc radius for their local measurements, while we use 1\,kpc. 

\begin{table*}[!htbp]
    \centering
    \movetableright=-1in
    \begin{tabular}{|c|c|c|c|c|c|}
        \hline
         Quantity & Sample Size & Median & Step Size (mag) & Significance & Tension\\ 
         \hline
         Local $U-V$          & 27 & 0.584 & -0.085 $\pm$ 0.086 & 1.0$\sigma$  & 0.35$\sigma$ \\
         Global $U-V$         & 27 & 0.646 & -0.062 $\pm$ 0.084 & 0.65$\sigma$ & 0.56$\sigma$ \\
         Host Stellar Mass    & 27 & 10.7  & -0.056 $\pm$ 0.086 & 0.65$\sigma$ & 0.33$\sigma$ \\
        \hline
    \end{tabular}
    \caption{Measurements of step size and significance for steps in local $U-V$ color, global $U-V$ color, and mass as measured in \citet{roman2018}. Here ``Tension" is the tension with the measurement in \citet{roman2018} of Ia SNe with $z<0.1$.}
    \label{tab:roman}
\end{table*} 

Our constraints are not in statistical tension with those found in \citet{roman2018}, as would be expected for a subsample of this size. We obtain a constraint on a host-galaxy stellar mass step of $-0.056 \pm$ 0.086\,mag for the sample of 27 overlapping galaxies. We find a step of $-0.085 \pm 0.086$\,mag for the local $U-V$ color, measured in a radius 3\,kpc around the SN. However, we note that our small sample size limits our interpretation of these results. 

\section{Discussion and Conclusions} \label{sec:discussion}

We have found a step in Hubble residuals with respect to \logHa\ of $-0.097 \pm 0.051$ mag (1.9$\sigma$ significance), while \citet{rigault2020} find $0.045 \pm 0.029$ mag (1.6$\sigma$). The step we find is in the opposite direction from that found in the \citet{rigault2013} and \citet{rigault2020} measurements of local SFR. In our sample, SNe Ia in passive environments tend to be dimmer, on average, after light curve correction than their counterparts in star-forming environments. Comparing with \citet{rigault2020}, who also measured H$\alpha$ surface brightness, we find a tension of 2.4$\sigma$. Although our measurements of \logHa\ show strong correlation with those in \citet{rigault2013} and \citet{rigault2020}, the measurements are not consistent within their uncertainties. 

We note that our original sample of 82 SNe has only six SNe with HR$<-0.1$ and \logHa\ $<38.32$, which \cite{rigault2013} define as the ``$M_2$'' population. \citet{rigault2013} found that, among SN whose local environment had low \logHa, the distribution of HRs was bimodal. They labeled the SNe with relatively small HRs as $M_2$. In our sample, we find only six SNe that meet the criteria for $M_2$. This contrasts with seventeen $M_2$ SNe in \citet{rigault2013} out of a total of 82. Our sample without early-type galaxies has 82 SNe, so we would expect a comparable number of $M_2$ SNe, instead of the 6 we identify. After including twelve early-type galaxies, assuming the same fraction of $M_2$ SNe in the total sample, we would expect $\approx$19 $M_2$ SNe, as opposed to our eleven. 

In \citet{rigault2018}, the authors show that the $M_2$ population drives the \logHa\ step, so its absence would lead to the disappearance of the step, and perhaps even its reversal. We suspected that this might be due to sample selection, as our initial sample was selected only to include blue, star-forming galaxies. Among twelve early-type galaxies we found five SNe in our $M_2$ sample. However, the addition of these $M_2$ SNe only reduced the tension to 1.7$\sigma$, which is only modestly smaller than the 2.4$\sigma$ tension in our star-forming sample. %The reason why few SNe in our sample are $M_2$, even in galaxies with low SFR, is not evident. 

The \citet{rigault2013} sample of SNe consists primarily of SNe discovered by the galaxy-untargeted SN search. However, morphological classifications for the SN host galaxies have not been published.
When we use the host-galaxy morphological classifications from \citet{pruzh2020}, the fraction of early-type galaxies in our sample, after the addition of E and S0 targets, is comparable to that of the galaxy-untargeted SDSS survey. Of the 83 SDSS host galaxies classified by \citet{pruzh2020}, 16 are early type, or 19\%, so our proportion of early-type galaxies, 13\%, is comparable. As a test, we augmented the number of early-type galaxies in our sample by randomly selecting from our set until we reached a fraction of 19\% and recalculated the step. We repeated this process 100 times and found a median step of  $-0.022 \pm 0.045$ (0.48$\sigma$ significance), a tension of 1.2$\sigma$ with \citet{rigault2020}. The exercise reduces but does not eliminate the tension. We conclude that while sample selection may play some role, a much larger proportion of early-type galaxies than found in low-redshift, galaxy-untargeted surveys such as SDSS would be necessary to eliminate the tension.

The addition of early-type galaxies may also speak to the question of how important the local environment is in standardizing SNe Ia versus the global environment. The early-type galaxies were selected solely based on their global properties. Although we performed local surface-brightness measurements for these galaxies as well, they all fell into the low \logHa group, as would be expected from their global properties. Using the maximum likelihood method (with a step) to fit the HRs for all early-type galaxies, we find that the best fit for the mean value is $-0.041 \pm 0.14$. For spiral-type galaxies, the best fit for the mean value of the HRs is $0.029 \pm 0.052$. The statistical tension between the HRs for the SNe in early-type and spiral galaxies has a statistical significance of only 0.46$\sigma$, so there is no significant evidence that galaxy morphology, a global property of the host, is responsible for the patterns we observe.

Both our sample and the \citet{rigault2020} distances use the SALT2 light curve fitter. However, our sample consists of SNe found almost entirely by targeted surveys, while the SNFactory sample used by \citet{rigault2013} and \citet{rigault2020} is largely from untargeted surveys. 
An untargeted survey is more likely to find SNe in less massive galaxies, and the $M_2$ population may potentially be connected to passive, low-mass galaxies. 
The SNe in our sample are also at lower redshifts, with our median redshift of $z=0.026$ being less than the $z=0.03$ lower limit of the \citet{rigault2013} sample. The difference between the redshift distributions could possibly yield a selection effect. 

\citet{roman2018} and \citet{jones2018} have performed the most comparable analyses to our study. \citet{roman2018} find that their results are compatible with \citet{rigault2013}, while \citet{jones2018} are not in full agreement. \citet{jones2018} suggest that their tension may be the result of targeted versus untargeted surveys, which our work supports as well. While \citet{roman2018} include SNe discovered by targeted surveys, their survey extends beyond redshift $z=0.5$, and therefore redshift evolution may become a factor. \citet{roman2018} and \citet{jones2018} also both use SED fitting to photometry to infer a SFR. The broadband UV-through-optical SED has sensitivity to star formation over a much longer period ($\sim$100 Myrs) than the luminosity of nebular H$\alpha$ emission, which requires massive O and B type stars ($\lesssim$10 Myrs). Calibration of SN Ia luminosities depends upon  the properties of host-galaxy dust, including its $R_V$ parameter. Since dust creation and destruction has connections with massive star formation through stellar winds and core-collapse SNe, the properties of the dust associated with high H$\alpha$ surface brightness could also differ potentially from those in regions with even somewhat less recent star formation.

We are not able to explain the statistical tension between our results and those presented by \citet{rigault2013} and \citet{rigault2020}. In later papers in this series, we plan to continue to use our sample of IFU spectra to investigate correlations between additional host galaxy properties, including metallicity, and HRs. We will also investigate the role of dust. Discoveries from the ZTF \citep[e.g.,][]{rigault2024, ginolin2024a,ginolin2024b} as well as future surveys such as the Legacy Survey of Space and Time with the Vera C. Rubin telescope will provide large untargeted samples that will allow for a more comprehensive comparison. 

\section{Acknowledgements}

We thank Coyne Gibson, Illa Rivero Losada, and David Doss for observing support and the staff at McDonald Observatory and the Harlan J. Smith telescope. We thank Hayley Williams for assistance in writing code for this project. We also thank our anonymous referee for their helpful comments.

A.M.I. and P.L.K are supported by NASA ADAP program 80NSSC21K0992. P.L.K. also acknowledges NSF AAG program AST-2308051. J.C.W. was supported in part by NSF grant 1813825. 

The Pan-STARRS1 Surveys (PS1) and the PS1 public science archive have been made possible through contributions by the Institute for Astronomy, the University of Hawaii, the Pan-STARRS Project Office, the Max-Planck Society and its participating institutes, the Max Planck Institute for Astronomy, Heidelberg and the Max Planck Institute for Extraterrestrial Physics, Garching, The Johns Hopkins University, Durham University, the University of Edinburgh, the Queen's University Belfast, the Harvard-Smithsonian Center for Astrophysics, the Las Cumbres Observatory Global Telescope Network Incorporated, the National Central University of Taiwan, the Space Telescope Science Institute, the National Aeronautics and Space Administration under Grant No. NNX08AR22G issued through the Planetary Science Division of the NASA Science Mission Directorate, the National Science Foundation Grant No. AST-1238877, the University of Maryland, Eotvos Lorand University (ELTE), the Los Alamos National Laboratory, and the Gordon and Betty Moore Foundation. 

The reference images in this paper that are from PS1 and can be accessed via \dataset[https://doi.org/10.17909/s0zg-jx37]{https://doi.org/10.17909/s0zg-jx37}. These images were retrieved from the Mikulski Archive for Space Telescopes (MAST) at the Space Telescope Science Institute. STScI is operated by the Association of Universities for Research in Astronomy, Inc., under NASA contract NAS5–26555. Support to MAST for these data is provided by the NASA Office of Space Science via grant NAG5–7584 and by other grants and contracts.

This work is in part based on observations obtained with the Samuel Oschin Telescope 48-inch and the 60-inch Telescope at the Palomar Observatory as part of the Zwicky Transient Facility project. ZTF is supported by the National Science Foundation under Grant No. AST-2034437 and a collaboration including Caltech, IPAC, the Weizmann Institute for Science, the Oskar Klein Center at Stockholm University, the University of Maryland, Deutsches Elektronen-Synchrotron and Humboldt University, the TANGO Consortium of Taiwan, the University of Wisconsin at Milwaukee, Trinity College Dublin, Lawrence Livermore National Laboratories, and IN2P3, France. Operations are conducted by COO, IPAC, and UW.

Funding for the Sloan Digital Sky Survey IV has been provided by the Alfred P. Sloan Foundation, the U.S. Department of Energy Office of Science, and the Participating Institutions. 

SDSS-IV acknowledges support and resources from the Center for High Performance Computing at the University of Utah. The SDSS website is www.sdss4.org.

SDSS-IV is managed by the Astrophysical Research Consortium for the Participating Institutions of the SDSS Collaboration including the Brazilian Participation Group, the Carnegie Institution for Science, Carnegie Mellon University, Center for Astrophysics | Harvard \& Smithsonian, the Chilean Participation Group, the French Participation Group, Instituto de Astrof\'isica de Canarias, The Johns Hopkins University, Kavli Institute for the Physics and Mathematics of the Universe (IPMU) / University of Tokyo, the Korean Participation Group, Lawrence Berkeley National Laboratory, Leibniz Institut f\"ur Astrophysik Potsdam (AIP),  Max-Planck-Institut f\"ur Astronomie (MPIA Heidelberg), Max-Planck-Institut f\"ur Astrophysik (MPA Garching), Max-Planck-Institut f\"ur Extraterrestrische Physik (MPE), National Astronomical Observatories of China, New Mexico State University, New York University, University of Notre Dame, Observat\'ario Nacional / MCTI, The Ohio State University, Pennsylvania State University, Shanghai Astronomical Observatory, United Kingdom Participation Group, Universidad Nacional Aut\'onoma de M\'exico, University of Arizona, University of Colorado Boulder, University of Oxford, University of Portsmouth, University of Utah, University of Virginia, University of Washington, University of Wisconsin, Vanderbilt University, and Yale University.

Funding for the SDSS and SDSS-II has been provided by the Alfred P. Sloan Foundation, the Participating Institutions, the National Science Foundation, the U.S. Department of Energy, the National Aeronautics and Space Administration, the Japanese Monbukagakusho, the Max Planck Society, and the Higher Education Funding Council for England. The SDSS Web Site is http://www.sdss.org/.

The SDSS is managed by the Astrophysical Research Consortium for the Participating Institutions. The Participating Institutions are the American Museum of Natural History, Astrophysical Institute Potsdam, University of Basel, University of Cambridge, Case Western Reserve University, University of Chicago, Drexel University, Fermilab, the Institute for Advanced Study, the Japan Participation Group, Johns Hopkins University, the Joint Institute for Nuclear Astrophysics, the Kavli Institute for Particle Astrophysics and Cosmology, the Korean Scientist Group, the Chinese Academy of Sciences (LAMOST), Los Alamos National Laboratory, the Max-Planck-Institute for Astronomy (MPIA), the Max-Planck-Institute for Astrophysics (MPA), New Mexico State University, Ohio State University, University of Pittsburgh, University of Portsmouth, Princeton University, the United States Naval Observatory, and the University of Washington.

\bibliography{Halpha_Bib}
\bibliographystyle{aasjournal}

\appendix 

\section{Weighted Average Method} \label{app1}
We used the inverse of the variance as the weight when taking the weighted average on either side of the step. The uncertainty of the mean is then given by, 

\begin{equation}
    \sigma=\frac{1}{\sum(\frac{1}{\sigma_i^2})}.
    \label{eqn:ave}
\end{equation}

This method does not take uncertainty on independent variable into account but more closely resembles the method used in \citet{rigault2013}. We feel that the ability to account for all uncertainty is valuable to our analysis, so we used the Monte Carlo method for our main results. The tables and figures presented here reflect the conclusions drawn from the weighted-average method. We used same populations for both calculations. 

Table \ref{app:Ha} shows the results for the star-forming sample. Table \ref{app:ellipticals} shows the results when early-type galaxies are added. We find a tension with the result from \citet{rigault2020} of 3.1$\sigma$ for the star-forming sample and 2.3$\sigma$ for the sample with added early-type galaxies, greater than the tensions from the MC results. As above, adding early-type galaxies reduces the tension with \citet{rigault2020} but does not eliminate it. 

We find that the trend towards higher steps with the median from \citet{rigault2013} holds, as does the trend towards larger steps after BPT cuts. However, unlike with the MC method, there is no consistent trend when performing light curve cuts.  

\begin{table}[!htbp]
\movetableright=-0.2in
    \begin{tabular}{|c|c|c|c|c|}
        \hline
        \multicolumn{5}{|c|}{All SNe} \\
        \hline
         Step Location (Log$_{10}$($\Sigma_{H\alpha}$ [erg s$^{-1}$ kpc$^{-2}$])) & Cuts & Sample Size & Step Size (mag) & Significance \\ \hline
         38.30 (This work)         & ... & 82 & -0.075 $\pm$ 0.045 & 1.6$\sigma$ \\
         38.46 (This work)         & BPT & 47 & -0.091 $\pm$ 0.061 & 1.5$\sigma$ \\
         38.32 \citep{rigault2013} & ... & 82 & -0.063 $\pm$ 0.045 & 1.4$\sigma$ \\
         38.32 \citep{rigault2013} & BPT & 47 & -0.13  $\pm$ 0.072 & 1.7$\sigma$ \\
        \hline
        \multicolumn{5}{|c|}{Light Curve Cuts} \\
        \hline
         Step Location (Log$_{10}$($\Sigma_{H\alpha}$ [erg s$^{-1}$ kpc$^{-2}$])) & Cuts & Sample Size & Step Size (mag) & Significance \\ \hline
         38.33 (This work)         & ... & 73 & -0.092 $\pm$ 0.048 & 1.9$\sigma$ \\
         38.46 (This work)         & BPT & 45 & -0.11  $\pm$ 0.062 & 1.7$\sigma$ \\
         38.32 \citep{rigault2013} & ... & 73 & -0.093 $\pm$ 0.048 & 2.0$\sigma$ \\
         38.32 \citep{rigault2013} & BPT & 45 & \-0.15  $\pm$ 0.073 & 2.0$\sigma$ \\
        \hline
    \end{tabular}
    \caption{Measurements of step size and significance for \logHa\ step. ``BPT" refers to the cuts based on the BPT diagram. ``Light Curve Cuts" refers to the light curve cuts based on color and stretch detailed in Sec.~\ref{sec:analysis}. The first two lines in each section show the results when the step location is the median of the data; the next two lines show the results when using the location used in \cite{rigault2013}.}
    \label{app:Ha}
\end{table} 

\begin{table}[!htbp]
    \centering
    \movetableright=-0.7in
    \begin{tabular}{|c|c|c|c|c|}
        \hline
        \multicolumn{5}{|c|}{All SNe} \\
        \hline
         Step Location (Log$_{10}$($\Sigma_{H\alpha}$ [erg s$^{-1}$ kpc$^{-2}$])) & Cuts & Sample Size & Step Size (mag) & Significance \\ \hline
         38.13 (This work)         & None & 94 & -0.021 $\pm$ 0.042 & 0.51$\sigma$ \\
         38.46 (This work)         & BPT  & 48 & -0.095 $\pm$ 0.061 & 1.6$\sigma$ \\
         38.32 \citep{rigault2013} & None & 94 & -0.037 $\pm$ 0.042 & 0.87$\sigma$ \\
         38.32 \citep{rigault2013} & BPT  & 48 & -0.13  $\pm$ 0.072 & 1.8$\sigma$ \\
        \hline
        \multicolumn{5}{|c|}{Light Curve Cuts} \\
        \hline
         Step Location (Log$_{10}$($\Sigma_{H\alpha}$ [erg s$^{-1}$ kpc$^{-2}$])) & Cuts & Sample Size & Step Size (mag) & Significance \\ \hline
         38.27 (This work)         & None & 81 & -0.058 $\pm$ 0.045 & 1.3$\sigma$ \\
         38.46 (This work)         & BPT  & 45 & -0.11  $\pm$ 0.062 & 1.7$\sigma$ \\
         38.32 \citep{rigault2013} & None & 81 & -0.050 $\pm$ 0.045 & 1.1$\sigma$ \\
         38.32 \citep{rigault2013} & BPT  & 45 & -0.15  $\pm$ 0.073 & 2.0$\sigma$ \\
        \hline
    \end{tabular}
    \caption{Measurements of step size and significance for \logHa\ step with added early-type galaxies. ``BPT" refers to the cuts based on the BPT diagram. ``Light Curve Cuts" refers to the light curve cuts based on color and stretch detailed in Sec.~\ref{sec:analysis}. The first two lines in each section show the results when the step location is the median of the data; the next two lines show the results when using the location used in \cite{rigault2013}.}
    \label{app:ellipticals}
\end{table} 

With the 42 SNe present in both this sample and that of \cite{betoule2014}, we find steps of $-0.069 \pm 0.062$ with Pantheon+ and $-0.052 \pm 0.051$ with \citet{betoule2014}. The tension between the measurements is 0.21$\sigma$, meaning that the values are consistent with each other at the 1$\sigma$ level. With the 49 SNe in both this sample and that of \citet{hicken2009}, we find steps of $-0.058 \pm 0.058$ with Pantheon+ and $-0.026 \pm 0.058$ with \citet{hicken2009}. The tension between these measurements is 0.39$\sigma$, again showing that the measurements are consistent with each other at the 1$\sigma$ level. 

The equivalent width calculations using the weighted average method are shown in Table \ref{app:EW}. We again find no correlation between H$\alpha$ EW and HRs.  

\begin{table}[!htbp]
    \centering
    \movetableright=-0.7in
    \begin{tabular}{|c|c|c|c|c|}
        \hline
        \multicolumn{5}{|c|}{Host Galaxies Selected with Star-Forming Morphology} \\
        \hline
         Step Location (\AA) & Cuts & Sample Size & Step Size (mag) & Significance \\ \hline
         -9.30 & ...     & 82 & 0.0094  $\pm$ 0.045 & 0.21$\sigma$ \\
         -9.45 & LC      & 73 & 0.0050  $\pm$ 0.048 & 0.10$\sigma$ \\
         -12.0 & BPT     & 47 & -0.0043 $\pm$ 0.061 & 0.070$\sigma$ \\
         -11.3 & BPT, LC & 45 & -0.032  $\pm$ 0.062 & 0.52$\sigma$ \\
        \hline
        \multicolumn{5}{|c|}{With Early-type Galaxies} \\
        \hline
         Step Location (\AA) & Cuts & Sample Size & Step Size (mag) & Significance \\ \hline
         -7.73 & ...     & 94 & 0.00075 $\pm$ 0.042 & 0.018$\sigma$ \\
         -7.99 & LC      & 81 & 0.027   $\pm$ 0.045 & 0.61$\sigma$ \\
         -11.6 & BPT     & 48 & -0.0086 $\pm$ 0.061 & 0.14$\sigma$ \\
         -11.3 & BPT, LC & 45 & -0.032  $\pm$ 0.062 & 0.52$\sigma$ \\
         \hline
    \end{tabular}
    \caption{Measurements of step size and significance for Halpha equivalent width. ``BPT" refers to the cuts using the \citet{kewley2001} line to exclude AGN. ``LC" refers to light curve cuts based on color stretch as detailed in Sec.~\ref{sec:analysis}. The upper section shows the results before the addition of early-type galaxies, and the lower table shows the results after.}
    \label{app:EW}
\end{table} 

We also repeated our analysis of the data from \citet{roman2018}, shown in Table \ref{app:roman}. We again find that our results are not in tension with those from \citet{roman2018}, that the mass step and global color steps are both insignificant, and the local $U-V$ color shows a mild effect. 

\begin{table}
    \movetableright=-0.8in
    \centering
    \begin{tabular}{|c|c|c|c|c|}
        \hline
         Quantity & Median & Step Size (mag) & Significance & Tension\\ 
         \hline
         Local $U-V$  & 0.584 & -0.073 $\pm$ 0.055 & 1.3$\sigma$  & 0.31$\sigma$ \\
         Global $U-V$ & 0.646 & -0.019 $\pm$ 0.055 & 0.35$\sigma$ & 0.12$\sigma$ \\
         Host Mass    & 10.7  & -0.018 $\pm$ 0.055 & 0.33$\sigma$ & 0.12$\sigma$ \\
        \hline
    \end{tabular}
    \caption{Measurements of step size and significance for steps in local $U-V$ color, global $U-V$ color, and host mass as measured in \citet{roman2018}. Here ``Tension" is the tension with the measurement in \citet{roman2018} of Ia SNe with $z<0.1$.}
    \label{app:roman}
\end{table} 

\pagebreak

\section{Full Data}

\startlongtable
\begin{deluxetable}{|l|l|l|l|l|l|}
    \tablehead{
    \colhead{Object} & \colhead{RA} & \colhead{DEC} & \colhead{z\tablenotemark{a}} & \colhead{Distance (Mpc)} & \colhead{Extraction Radius}}
    \centering
    \startdata
        SN1990O & 258.9 & 16.32 & 0.031$\pm$ 0.0012 & 127.6$\pm$7.43 & 1.62$\pm$ 0.083\\
        SN1994M & 187.79 & 0.61 & 0.023$\pm$ 0.0012 & 100.94$\pm$7.32 & 2.04$\pm$ 0.148\\
        SN1994S & 187.84 & 29.13 & 0.016$\pm$ 0.0012 & 67.99$\pm$7.24 & 3.03$\pm$ 0.303\\
        SN1995ac & 341.39 & -8.75 & 0.049$\pm$ 0.0012 & 196.58$\pm$7.59 & 1.05$\pm$ 0.034\\
        SN1996C & 207.7 & 49.32 & 0.029$\pm$ 0.0012 & 118.91$\pm$7.38 & 1.73$\pm$ 0.096\\
        SN1996bl & 9.07 & 11.39 & 0.035$\pm$ 0.0012 & 142.94$\pm$7.46 & 1.44$\pm$ 0.066\\
        SN1996bv & 94.05 & 57.05 & 0.017$\pm$ 0.0012 & 70.21$\pm$7.25 & 2.94$\pm$ 0.284\\
        SN1997do & 111.68 & 47.09 & 0.01$\pm$ 0.0012 & 44.31$\pm$7.18 & 4.65$\pm$ 0.724\\
        SN1998V & 275.66 & 15.7 & 0.017$\pm$ 0.0012 & 72.06$\pm$7.25 & 2.86$\pm$ 0.269\\
        SN1998ab & 192.2 & 41.92 & 0.028$\pm$ 0.0012 & 116.16$\pm$7.37 & 1.78$\pm$ 0.101\\
        SN1998dx & 272.8 & 49.86 & 0.05$\pm$ 0.0012 & 200.88$\pm$7.60 & 1.03$\pm$ 0.032\\
        SN1998es & 24.32 & 5.88 & 0.01$\pm$ 0.0012 & 40.51$\pm$7.23 & 5.09$\pm$ 0.875\\
        SN1999aa & 126.93 & 21.49 & 0.016$\pm$ 0.0012 & 66.34$\pm$7.24 & 3.11$\pm$ 0.319\\
        SN1999cc & 240.68 & 37.36 & 0.031$\pm$ 0.0012 & 129.15$\pm$7.40 & 1.6$\pm$ 0.081\\
        SN1999dg & 227.87 & 13.48 & 0.023$\pm$ 0.0012 & 94.29$\pm$7.36 & 2.19$\pm$ 0.156\\
        SN1999dk & 22.86 & 14.28 & 0.014$\pm$ 0.0012 & 58.95$\pm$7.22 & 3.5$\pm$ 0.405\\
        SN1999dq & 38.5 & 20.98 & 0.013$\pm$ 0.0012 & 56.21$\pm$7.21 & 3.67$\pm$ 0.447\\
        SN1999gp & 37.91 & 39.38 & 0.026$\pm$ 0.0012 & 108.02$\pm$7.35 & 1.91$\pm$ 0.117\\
        SN2000cf & 238.23 & 65.94 & 0.036$\pm$ 0.0012 & 149.42$\pm$7.47 & 1.38$\pm$ 0.060\\
        SN2000cw & 356.85 & 28.39 & 0.029$\pm$ 0.0012 & 119.75$\pm$7.38 & 1.72$\pm$ 0.095\\
        SN2000dg & 1.56 & 8.89 & 0.037$\pm$ 0.0012 & 152.51$\pm$7.47 & 1.35$\pm$ 0.057\\
        SN2000dk & 16.85 & 32.41 & 0.016$\pm$ 0.0012 & 69.22$\pm$7.27 & 2.98$\pm$ 0.293\\
        SN2000dn & 346.27 & -3.2 & 0.031$\pm$ 0.0012 & 127.12$\pm$7.53 & 1.62$\pm$ 0.085\\
        SN2001ah & 167.62 & 55.16 & 0.058$\pm$ 0.0012 & 233.08$\pm$7.69 & 0.88$\pm$ 0.023\\
        SN2001az & 248.62 & 76.03 & 0.041$\pm$ 0.0012 & 165.53$\pm$7.52 & 1.25$\pm$ 0.048\\
        SN2001cj & 200.44 & 31.25 & 0.025$\pm$ 0.0012 & 103.51$\pm$7.34 & 1.99$\pm$ 0.128\\
        SN2001ck & 219.46 & 30.48 & 0.035$\pm$ 0.0012 & 145.42$\pm$7.79 & 1.42$\pm$ 0.066\\
        SN2001da & 358.39 & 8.12 & 0.017$\pm$ 0.0012 & 69.59$\pm$7.25 & 2.96$\pm$ 0.289\\
        SN2001dl & 320.26 & 9.18 & 0.021$\pm$ 0.0012 & 82.9$\pm$7.30 & 2.21$\pm$ 0.172\\
        SN2001eh & 24.55 & 41.66 & 0.036$\pm$ 0.0012 & 148.67$\pm$7.53 & 1.39$\pm$ 0.061\\
        SN2001en & 21.35 & 34.03 & 0.015$\pm$ 0.0012 & 64.48$\pm$7.23 & 3.2$\pm$ 0.338\\
        SN2001fe & 144.49 & 25.49 & 0.014$\pm$ 0.0012 & 60.93$\pm$7.22 & 3.39$\pm$ 0.379\\
        SN2002de & 244.13 & 35.71 & 0.028$\pm$ 0.0012 & 125.89$\pm$7.37 & 1.69$\pm$ 0.102\\
        SN2002hu & 34.58 & 37.47 & 0.036$\pm$ 0.0015 & 147.26$\pm$9.82 & 1.4$\pm$ 0.081\\
        SN2002jy & 20.32 & 40.5 & 0.02$\pm$ 0.0012 & 78.29$\pm$7.29 & 2.34$\pm$ 0.194\\
        SN2003U & 260.69 & 62.16 & 0.028$\pm$ 0.0012 & 116.97$\pm$7.37 & 1.66$\pm$ 0.099\\
        SN2003ch & 109.49 & 9.69 & 0.029$\pm$ 0.0012 & 127.66$\pm$7.42 & 1.62$\pm$ 0.094\\
        SN2003fa & 266.03 & 40.88 & 0.04$\pm$ 0.0012 & 163.82$\pm$7.54 & 1.26$\pm$ 0.050\\
        SN2003gn & 338.47 & 20.8 & 0.033$\pm$ 0.0012 & 136.43$\pm$7.42 & 1.51$\pm$ 0.072\\
        SN2003ic & 10.46 & -9.31 & 0.052$\pm$ 0.0012 & 209.06$\pm$7.62 & 0.99$\pm$ 0.029\\
        SN2003it & 1.45 & 27.45 & 0.024$\pm$ 0.0012 & 100.36$\pm$7.33 & 2.06$\pm$ 0.136\\
        SN2003iv & 42.53 & 12.85 & 0.034$\pm$ 0.0012 & 149.88$\pm$7.50 & 1.38$\pm$ 0.069\\
        SN2004at & 164.69 & 59.49 & 0.023$\pm$ 0.0012 & 95.26$\pm$7.31 & 2.17$\pm$ 0.152\\
        SN2004bg & 170.26 & 21.34 & 0.022$\pm$ 0.0012 & 92.31$\pm$7.31 & 2.23$\pm$ 0.162\\
        SN2004bk & 204.37 & 4.1 & 0.024$\pm$ 0.0012 & 100.04$\pm$7.33 & 2.06$\pm$ 0.137\\
        SN2004br & 187.78 & 0.61 & 0.024$\pm$ 0.0012 & 100.89$\pm$7.33 & 2.04$\pm$ 0.135\\
        SN2004ef & 340.54 & 19.99 & 0.03$\pm$ 0.0012 & 123.02$\pm$7.39 & 1.68$\pm$ 0.090\\
        SN2005M & 144.38 & 23.2 & 0.026$\pm$ 0.0012 & 107.25$\pm$7.35 & 1.92$\pm$ 0.119\\
        SN2005bg & 184.32 & 16.37 & 0.024$\pm$ 0.0012 & 100.4$\pm$7.33 & 2.05$\pm$ 0.136\\
        SN2005de & 270.6 & 26.05 & 0.015$\pm$ 0.0012 & 62.87$\pm$7.24 & 3.28$\pm$ 0.356\\
        SN2005eu & 36.93 & 28.18 & 0.034$\pm$ 0.0051 & 138.52$\pm$32.70 & 1.49$\pm$ 0.308\\
        SN2005hc & 29.2 & -0.21 & 0.045$\pm$ 0.0012 & 182.4$\pm$7.55 & 1.13$\pm$ 0.039\\
        SN2005ms & 132.31 & 36.13 & 0.026$\pm$ 0.0012 & 107.69$\pm$7.35 & 1.92$\pm$ 0.118\\
        SN2005na & 105.4 & 14.13 & 0.027$\pm$ 0.0012 & 112.59$\pm$7.38 & 1.83$\pm$ 0.108\\
        SN2006N & 92.13 & 64.72 & 0.014$\pm$ 0.0012 & 59.61$\pm$7.24 & 3.46$\pm$ 0.397\\
        SN2006S & 191.41 & 35.09 & 0.033$\pm$ 0.0012 & 136.19$\pm$7.42 & 1.51$\pm$ 0.073\\
        SN2006bw & 218.49 & 3.8 & 0.031$\pm$ 0.0012 & 126.88$\pm$7.40 & 1.63$\pm$ 0.084\\
        SN2006cf & 163.51 & 46.03 & 0.042$\pm$ 0.0012 & 171.88$\pm$7.52 & 1.2$\pm$ 0.045\\
        SN2006cp & 184.81 & 22.43 & 0.023$\pm$ 0.0012 & 97.17$\pm$7.32 & 2.12$\pm$ 0.146\\
        SN2006et & 10.69 & -23.56 & 0.021$\pm$ 0.0012 & 89.3$\pm$7.36 & 2.31$\pm$ 0.175\\
        SN2006gr & 338.09 & 30.83 & 0.034$\pm$ 0.0012 & 137.85$\pm$7.43 & 1.5$\pm$ 0.071\\
        SN2006ot & 33.77 & -20.77 & 0.052$\pm$ 0.0012 & 210.42$\pm$7.69 & 0.98$\pm$ 0.029\\
        SN2006py & 340.43 & -0.14 & 0.057$\pm$ 0.0012 & 226.41$\pm$7.67 & 0.91$\pm$ 0.025\\
        SN2006sr & 0.9 & 23.2 & 0.023$\pm$ 0.0012 & 96.07$\pm$7.32 & 2.15$\pm$ 0.149\\
        SN2007A & 6.32 & 12.89 & 0.017$\pm$ 0.0012 & 75.48$\pm$7.28 & 2.73$\pm$ 0.264\\
        SN2007F & 195.81 & 50.62 & 0.024$\pm$ 0.0012 & 100.64$\pm$7.33 & 2.05$\pm$ 0.136\\
        SN2007O & 224.02 & 45.4 & 0.037$\pm$ 0.0012 & 152.79$\pm$7.47 & 1.35$\pm$ 0.057\\
        SN2007ai & 243.22 & -21.63 & 0.032$\pm$ 0.0012 & 132.32$\pm$7.47 & 1.56$\pm$ 0.078\\
        SN2007bc & 169.81 & 20.81 & 0.022$\pm$ 0.0012 & 91.21$\pm$7.30 & 2.26$\pm$ 0.166\\
        SN2007ca & 202.77 & -15.1 & 0.015$\pm$ 0.0012 & 63.45$\pm$7.23 & 3.25$\pm$ 0.349\\
        SN2007co & 275.77 & 29.9 & 0.027$\pm$ 0.0012 & 110.11$\pm$7.39 & 1.87$\pm$ 0.113\\
        SN2007cq & 333.67 & 5.08 & 0.025$\pm$ 0.0012 & 103.1$\pm$7.35 & 2$\pm$ 0.129\\
        SN2007is & 251.81 & 40.24 & 0.03$\pm$ 0.0012 & 122.03$\pm$7.39 & 1.69$\pm$ 0.091\\
        SN2007jh & 54.01 & 1.1 & 0.04$\pm$ 0.0012 & 164.13$\pm$7.50 & 1.26$\pm$ 0.049\\
        SN2007kk & 55.6 & 39.24 & 0.041$\pm$ 0.0012 & 168.28$\pm$7.58 & 1.23$\pm$ 0.047\\
        SN2007qe & 358.55 & 27.41 & 0.023$\pm$ 0.0012 & 94.98$\pm$7.32 & 2.17$\pm$ 0.153\\
        SN2007sw & 183.4 & 46.49 & 0.025$\pm$ 0.0012 & 103.87$\pm$7.35 & 1.99$\pm$ 0.127\\
        SN2008C & 104.3 & 20.44 & 0.017$\pm$ 0.0012 & 71.69$\pm$7.25 & 2.88$\pm$ 0.272\\
        SN2008Z & 145.81 & 36.28 & 0.021$\pm$ 0.0012 & 89.54$\pm$7.30 & 2.3$\pm$ 0.173\\
        SN2008ar & 186.16 & 10.84 & 0.027$\pm$ 0.0012 & 113.19$\pm$7.36 & 1.82$\pm$ 0.106\\
        SN2008bf & 181.01 & 20.25 & 0.025$\pm$ 0.0015 & 101.89$\pm$9.62 & 2.02$\pm$ 0.173\\
        SN2008gb & 44.49 & 46.87 & 0.036$\pm$ 0.0012 & 149.22$\pm$7.48 & 1.38$\pm$ 0.060\\
        SN2008gp & 50.75 & 1.36 & 0.032$\pm$ 0.0012 & 133.62$\pm$7.52 & 1.54$\pm$ 0.077\\
        SN2008hj & 1.01 & -11.17 & 0.036$\pm$ 0.0012 & 149.3$\pm$7.52 & 1.38$\pm$ 0.060\\
        SN2009D & 58.6 & -19.18 & 0.025$\pm$ 0.0012 & 102.9$\pm$7.37 & 2$\pm$ 0.130\\
        SN2009ad & 75.89 & 6.66 & 0.028$\pm$ 0.0012 & 117.28$\pm$7.37 & 1.76$\pm$ 0.099\\
        SN2009dc & 237.8 & 25.71 & 0.022$\pm$ 0.0012 & 90.88$\pm$7.30 & 2.27$\pm$ 0.167\\
        SN2009ds & 177.27 & -9.73 & 0.02$\pm$ 0.0012 & 84.94$\pm$7.35 & 2.43$\pm$ 0.194\\
        SN2009kq & 129.06 & 28.07 & 0.012$\pm$ 0.0012 & 52.6$\pm$7.26 & 3.92$\pm$ 0.515\\
        SN2009na & 161.76 & 26.54 & 0.022$\pm$ 0.0012 & 92.1$\pm$7.31 & 2.24$\pm$ 0.163\\
        SN2010Y & 162.77 & 65.78 & 0.011$\pm$ 0.0012 & 48.23$\pm$7.19 & 4.28$\pm$ 0.609\\
        SN2010dt & 250.81 & 32.68 & 0.053$\pm$ 0.0012 & 212.64$\pm$7.63 & 0.97$\pm$ 0.028\\
        SN2010kg & 70.04 & 7.35 & 0.016$\pm$ 0.0012 & 67$\pm$7.24 & 3.08$\pm$ 0.312\\
        SNF20080514-002 & 202.3 & 11.27 & 0.023$\pm$ 0.0012 & 95.47$\pm$7.31 & 2.16$\pm$ 0.151\\
    \enddata
    \tablenotetext{a}{Includes additional uncertianty corresponding to 350 km/s peculiar velocity}
    \label{tab:data1}
\end{deluxetable}

\startlongtable
\begin{deluxetable}{|l|l|l|l|l|}
    \tablehead{
    \colhead{Object} & \colhead{$\mu_z$\tablenotemark{a}} & \colhead{$\mu_{SN}$\tablenotemark{b}} & \colhead{HR} & \colhead{\logHa}}
    \centering
    \startdata
        SN1990O & 35.53$\pm$0.119 & 35.56$\pm$0.14 & 0.03$\pm$0.184 & 37.6$\pm$0.0705\\
        SN1994M & 35.02$\pm$0.157 & 34.98$\pm$0.143 & -0.04$\pm$0.213 & 36$\pm$0.0009\\
        SN1994S & 34.16$\pm$0.224 & 34.1$\pm$0.169 & -0.058$\pm$0.281 & 38.07$\pm$0.0301\\
        SN1995ac & 36.47$\pm$0.076 & 36.4$\pm$0.125 & -0.071$\pm$0.146 & 38.26$\pm$0.0524\\
        SN1996C & 35.38$\pm$0.127 & 35.63$\pm$0.139 & 0.255$\pm$0.189 & 38.47$\pm$0.0327\\
        SN1996bl & 35.78$\pm$0.106 & 35.78$\pm$0.132 & 0.004$\pm$0.169 & 38.33$\pm$0.0441\\
        SN1996bv & 34.23$\pm$0.217 & 34.07$\pm$0.163 & -0.158$\pm$0.271 & 38.57$\pm$0.0239\\
        SN1997do & 33.23$\pm$0.345 & 33.25$\pm$0.209 & 0.018$\pm$0.403 & 39.09$\pm$0.0271\\
        SN1998V & 34.29$\pm$0.211 & 34.18$\pm$0.168 & -0.107$\pm$0.270 & 37.87$\pm$0.0346\\
        SN1998ab & 35.33$\pm$0.130 & 34.96$\pm$0.135 & -0.364$\pm$0.188 & 38.56$\pm$0.026\\
        SN1998dx & 36.51$\pm$0.075 & 36.66$\pm$0.134 & 0.143$\pm$0.153 & 36.89$\pm$0.4851\\
        SN1998es & 33.04$\pm$0.380 & 32.93$\pm$0.227 & -0.104$\pm$0.443 & 38.31$\pm$0.0242\\
        SN1999aa & 34.11$\pm$0.230 & 34.21$\pm$0.165 & 0.099$\pm$0.283 & 38.41$\pm$0.0391\\
        SN1999cc & 35.56$\pm$0.117 & 35.64$\pm$0.133 & 0.086$\pm$0.177 & 38.55$\pm$0.0388\\
        SN1999dg & 34.87$\pm$0.162 & 34.96$\pm$0.178 & 0.087$\pm$0.241 & 38.01$\pm$0.0463\\
        SN1999dk & 33.85$\pm$0.259 & 33.92$\pm$0.179 & 0.066$\pm$0.314 & 37.54$\pm$0.0636\\
        SN1999dq & 33.75$\pm$0.271 & 33.5$\pm$0.185 & -0.244$\pm$0.329 & 39.39$\pm$0.0402\\
        SN1999gp & 35.17$\pm$0.140 & 35.3$\pm$0.137 & 0.132$\pm$0.196 & 38.6$\pm$0.0233\\
        SN2000cf & 35.87$\pm$0.101 & 36.15$\pm$0.128 & 0.275$\pm$0.163 & 38.41$\pm$0.04\\
        SN2000cw & 35.39$\pm$0.126 & 35.71$\pm$0.161 & 0.315$\pm$0.205 & 37.67$\pm$0.0503\\
        SN2000dg & 35.92$\pm$0.099 & 36.19$\pm$0.204 & 0.269$\pm$0.227 & 38.18$\pm$0.0302\\
        SN2000dk & 34.2$\pm$0.221 & 34.19$\pm$0.161 & -0.011$\pm$0.273 & 37.42$\pm$0.0969\\
        SN2000dn & 35.52$\pm$0.121 & 35.81$\pm$0.162 & 0.287$\pm$0.202 & 36$\pm$0.0026\\
        SN2001ah & 36.84$\pm$0.064 & 37.04$\pm$0.172 & 0.204$\pm$0.183 & 36$\pm$0.0009\\
        SN2001az & 36.09$\pm$0.091 & 36.39$\pm$0.133 & 0.293$\pm$0.162 & 38.35$\pm$0.039\\
        SN2001cj & 35.07$\pm$0.147 & 35.31$\pm$0.163 & 0.236$\pm$0.219 & 36$\pm$0.0013\\
        SN2001ck & 35.81$\pm$0.109 & 35.96$\pm$0.162 & 0.146$\pm$0.195 & 38.39$\pm$0.0355\\
        SN2001da & 34.21$\pm$0.219 & 34.22$\pm$0.206 & 0.003$\pm$0.300 & 38.22$\pm$0.0289\\
        SN2001dl & 34.59$\pm$0.169 & 35.31$\pm$0.157 & 0.717$\pm$0.231 & 38.3$\pm$0.1507\\
        SN2001eh & 35.86$\pm$0.102 & 36.04$\pm$0.131 & 0.175$\pm$0.166 & 36$\pm$0.0009\\
        SN2001en & 34.05$\pm$0.236 & 33.99$\pm$0.17 & -0.061$\pm$0.291 & 38.26$\pm$0.0161\\
        SN2001fe & 33.92$\pm$0.250 & 33.98$\pm$0.174 & 0.057$\pm$0.305 & 38.68$\pm$0.0225\\
        SN2002de & 35.5$\pm$0.131 & 35.51$\pm$0.128 & 0.006$\pm$0.183 & 38.97$\pm$0.0089\\
        SN2002hu & 35.84$\pm$0.135 & 36.04$\pm$0.141 & 0.202$\pm$0.195 & 37$\pm$0.2738\\
        SN2002jy & 34.47$\pm$0.180 & 35.05$\pm$0.15 & 0.579$\pm$0.234 & 37.67$\pm$0.1832\\
        SN2003U & 35.32$\pm$0.129 & 35.32$\pm$0.146 & 0.008$\pm$0.195 & 38.09$\pm$0.0601\\
        SN2003ch & 35.39$\pm$0.126 & 35.66$\pm$0.139 & 0.269$\pm$0.188 & 37.22$\pm$0.1231\\
        SN2003fa & 36.07$\pm$0.092 & 36.14$\pm$0.126 & 0.066$\pm$0.156 & 36.92$\pm$0.767\\
        SN2003gn & 35.67$\pm$0.111 & 36.14$\pm$0.128 & 0.466$\pm$0.169 & 37.95$\pm$0.0456\\
        SN2003ic & 36.6$\pm$0.072 & 36.52$\pm$0.135 & -0.086$\pm$0.153 & 37.37$\pm$0.1397\\
        SN2003it & 35.01$\pm$0.151 & 35.11$\pm$0.147 & 0.1$\pm$0.211 & 38.06$\pm$0.0313\\
        SN2003iv & 35.7$\pm$0.109 & 35.98$\pm$0.139 & 0.288$\pm$0.176 & 36$\pm$0.0007\\
        SN2004at & 34.89$\pm$0.159 & 35.02$\pm$0.14 & 0.125$\pm$0.212 & 37.24$\pm$0.0767\\
        SN2004bg & 34.83$\pm$0.165 & 34.97$\pm$0.146 & 0.144$\pm$0.220 & 39.09$\pm$0.0186\\
        SN2004bk & 35$\pm$0.152 & 34.93$\pm$0.143 & -0.068$\pm$0.208 & 38.57$\pm$0.0377\\
        SN2004br & 35.02$\pm$0.150 & 34.91$\pm$0.14 & -0.109$\pm$0.205 & 36.83$\pm$0.6969\\
        SN2004ef & 35.45$\pm$0.123 & 35.5$\pm$0.146 & 0.047$\pm$0.191 & 37.95$\pm$0.0635\\
        SN2005M & 35.15$\pm$0.141 & 34.98$\pm$0.158 & -0.169$\pm$0.212 & 38.39$\pm$0.0249\\
        SN2005bg & 35.01$\pm$0.151 & 35.11$\pm$0.137 & 0.098$\pm$0.204 & 39.37$\pm$0.0387\\
        SN2005de & 33.99$\pm$0.243 & 34.29$\pm$0.166 & 0.297$\pm$0.294 & 37.11$\pm$0.0857\\
        SN2005eu & 35.71$\pm$0.480 & 35.91$\pm$0.339 & 0.199$\pm$0.588 & 38.14$\pm$0.0419\\
        SN2005hc & 36.31$\pm$0.082 & 36.57$\pm$0.125 & 0.265$\pm$0.150 & 37.92$\pm$0.0598\\
        SN2005ms & 35.16$\pm$0.141 & 35.4$\pm$0.137 & 0.243$\pm$0.197 & 36$\pm$0.0013\\
        SN2005na & 35.26$\pm$0.135 & 35.25$\pm$0.142 & -0.009$\pm$0.196 & 38.52$\pm$0.019\\
        SN2006N & 33.88$\pm$0.256 & 34$\pm$0.175 & 0.125$\pm$0.311 & 37.97$\pm$0.0432\\
        SN2006S & 35.67$\pm$0.111 & 35.94$\pm$0.13 & 0.27$\pm$0.171 & 38.37$\pm$0.0318\\
        SN2006bw & 35.52$\pm$0.119 & 35.56$\pm$0.14 & 0.047$\pm$0.184 & 36$\pm$0.0006\\
        SN2006cf & 36.18$\pm$0.087 & 36.38$\pm$0.133 & 0.204$\pm$0.159 & 38.65$\pm$0.027\\
        SN2006cp & 34.94$\pm$0.156 & 34.9$\pm$0.146 & -0.036$\pm$0.214 & 37.38$\pm$0.0807\\
        SN2006et & 34.75$\pm$0.172 & 34.77$\pm$0.164 & 0.013$\pm$0.237 & 38.3$\pm$0.0197\\
        SN2006gr & 35.7$\pm$0.110 & 35.95$\pm$0.13 & 0.252$\pm$0.170 & 36.43$\pm$1.0461\\
        SN2006ot & 36.62$\pm$0.072 & 36.53$\pm$0.177 & -0.082$\pm$0.191 & 37.11$\pm$0.757\\
        SN2006py & 36.77$\pm$0.066 & 37.01$\pm$0.117 & 0.239$\pm$0.134 & 37.44$\pm$0.108\\
        SN2006sr & 34.91$\pm$0.158 & 35.04$\pm$0.144 & 0.126$\pm$0.214 & 38.65$\pm$0.0407\\
        SN2007A & 34.2$\pm$0.209 & 34.55$\pm$0.166 & 0.354$\pm$0.267 & 39.04$\pm$0.024\\
        SN2007F & 35.01$\pm$0.151 & 35.18$\pm$0.14 & 0.165$\pm$0.206 & 38.23$\pm$0.0382\\
        SN2007O & 35.92$\pm$0.099 & 35.93$\pm$0.133 & 0.006$\pm$0.165 & 39.03$\pm$0.0299\\
        SN2007ai & 35.61$\pm$0.115 & 35.84$\pm$0.152 & 0.228$\pm$0.191 & 37.82$\pm$0.0545\\
        SN2007bc & 34.8$\pm$0.167 & 34.79$\pm$0.15 & -0.013$\pm$0.224 & 36$\pm$0.0012\\
        SN2007ca & 34.01$\pm$0.240 & 34.46$\pm$0.174 & 0.453$\pm$0.297 & 38.4$\pm$0.0263\\
        SN2007co & 35.21$\pm$0.138 & 35.34$\pm$0.137 & 0.133$\pm$0.195 & 38.42$\pm$0.0463\\
        SN2007cq & 35.07$\pm$0.147 & 34.94$\pm$0.138 & -0.125$\pm$0.202 & 38.27$\pm$0.0282\\
        SN2007is & 35.43$\pm$0.124 & 35.38$\pm$0.145 & -0.05$\pm$0.190 & 38.66$\pm$0.0238\\
        SN2007jh & 36.08$\pm$0.092 & 36.53$\pm$0.128 & 0.455$\pm$0.157 & 38.1$\pm$0.0615\\
        SN2007kk & 36.13$\pm$0.090 & 36.27$\pm$0.13 & 0.137$\pm$0.158 & 38.35$\pm$0.0469\\
        SN2007qe & 34.89$\pm$0.160 & 35.07$\pm$0.142 & 0.181$\pm$0.214 & 36$\pm$0.0008\\
        SN2007sw & 35.08$\pm$0.146 & 35.06$\pm$0.143 & -0.026$\pm$0.205 & 39.17$\pm$0.0192\\
        SN2008C & 34.28$\pm$0.212 & 34.28$\pm$0.161 & 0.007$\pm$0.267 & 39.23$\pm$0.0208\\
        SN2008Z & 34.76$\pm$0.170 & 35.35$\pm$0.147 & 0.586$\pm$0.225 & 37.57$\pm$0.0525\\
        SN2008ar & 35.27$\pm$0.134 & 35.46$\pm$0.135 & 0.187$\pm$0.190 & 38.5$\pm$0.0245\\
        SN2008bf & 35.04$\pm$0.195 & 34.95$\pm$0.165 & -0.093$\pm$0.256 & 36$\pm$0.0015\\
        SN2008gb & 35.87$\pm$0.101 & 36.19$\pm$0.141 & 0.324$\pm$0.174 & 38.36$\pm$0.0259\\
        SN2008gp & 35.63$\pm$0.115 & 35.77$\pm$0.129 & 0.141$\pm$0.173 & 37.12$\pm$0.1875\\
        SN2008hj & 35.87$\pm$0.102 & 35.99$\pm$0.143 & 0.119$\pm$0.175 & 37.59$\pm$0.0705\\
        SN2009D & 35.06$\pm$0.148 & 35.01$\pm$0.143 & -0.049$\pm$0.206 & 36.35$\pm$1.0018\\
        SN2009ad & 35.35$\pm$0.129 & 35.44$\pm$0.136 & 0.097$\pm$0.188 & 38.38$\pm$0.0456\\
        SN2009dc & 34.79$\pm$0.167 & 34.38$\pm$0.148 & -0.414$\pm$0.224 & 37.34$\pm$0.0677\\
        SN2009ds & 34.65$\pm$0.180 & 34.76$\pm$0.155 & 0.116$\pm$0.238 & 38.53$\pm$0.0378\\
        SN2009kq & 33.6$\pm$0.292 & 33.61$\pm$0.193 & 0.003$\pm$0.350 & 37.33$\pm$0.0561\\
        SN2009na & 34.82$\pm$0.165 & 34.84$\pm$0.148 & 0.019$\pm$0.221 & 38.73$\pm$0.0257\\
        SN2010Y & 33.42$\pm$0.316 & 33.79$\pm$0.193 & 0.377$\pm$0.371 & 38$\pm$0.0241\\
        SN2010dt & 36.64$\pm$0.070 & 36.93$\pm$0.145 & 0.288$\pm$0.161 & 37.45$\pm$0.1529\\
        SN2010kg & 34.13$\pm$0.227 & 34.01$\pm$0.202 & -0.121$\pm$0.304 & 38.51$\pm$0.0259\\
        SNF20080514-002 & 34.9$\pm$0.159 & 35.13$\pm$0.146 & 0.228$\pm$0.216 & 37.86$\pm$0.0322\\
    \enddata
    \tablenotetext{a}{Inferred from redshift and the cosmological parameters in \citet{brout2022}.}
    \tablenotetext{b}{From the Pantheon+ sample as given in \citet{brout2022}.}
    \label{tab:data2}
\end{deluxetable}

\end{document}